\documentclass[conference]{IEEEtran}

\IEEEoverridecommandlockouts

\usepackage{cite}
\usepackage{latexsym,theorem}
\usepackage[final]{fixme}
\usepackage[ruled]{algorithm2e}
\usepackage[utf8]{inputenc}
\usepackage{amssymb,
	    amsmath,
	    mathrsfs,
	    url,
	    color,
	    ifpdf,
	    setspace,
	    semantic,
	    xspace,
	    pstricks,
	    pst-node,
	    booktabs,
	    wrapfig,
	    colortbl}
\usepackage[final]{graphicx}
\usepackage{subfig}
\usepackage{hyperref}

\usepackage{amssymb}
\usepackage{xspace}
\usepackage{textcomp} 
\usepackage{calc}

\DeclareMathAlphabet\mathscra{T1}{hlcw}{m}{it}

\let\then\iftrue
{\catcode`\!=8 
 \catcode`\Q=3 
 \catcode`\@=11
 \long\gdef\ifgiven#1\then{\Ifbl@nk#1QQQ\empty!}
 \long\gdef\ifblank#1\then{\Ifbl@nk#1QQ..!}
 \long\gdef\Ifbl@nk#1#2Q#3!{\ifx#3}
 \long\gdef\ifnull#1\then{\IfN@LL#1* {#1}!}
 \long\gdef\IfN@LL#1 #2!{\ifblank{#2}\then}}

    {\bigskip\noindent\begin{tabular}{|p{\textwidth}}\it\small\strut}%
    {\strut\end{tabular}\par\bigskip}

    {\begingroup\vskip-2ex\begin{small}\noindent}%
    {\end{small}\vskip1.5mm\endgroup}

    {\begingroup\small\begin{equation}}%
    {\end{equation}\endgroup}
\newcommand*{\stignore}[1]{}

\newcommand{\wordsp}[1]{\textit{#1}}

\newcommand*{\mktype}[2]{%
  \expandafter\newcommand\csname #1sym\endcsname%
      {{\textsf{#2}}}%
  \expandafter\newcommand\csname #1name\endcsname%
      {{\csname #1sym\endcsname}\xspace}%
  \expandafter\newcommand\csname #1\endcsname%
      {{\csname #1sym\endcsname}\xspace}%
    }

\newcommand*{\mkkeyword}[2]{%
  \expandafter\newcommand\csname #1sym\endcsname%
      {{\ensuremath{#2}\expandafter\index{#2@\protect{\csname #1sym\endcsname}}}}%
  \expandafter\newcommand\csname #1name\endcsname%
      {{\csname #1sym\endcsname}\xspace}%
  \expandafter\newcommand\csname #1\endcsname%
      {{\csname #1sym\endcsname}\xspace}%
    }

\newcommand*{\mkfunc}[3]{%
  \expandafter\newcommand\csname #1sym\endcsname%
  {\ensuremath{#2}\index{#1@\protect{\csname #1sym\endcsname}}}%
  \expandafter\newcommand\csname #1name\endcsname%
  {\csname #1sym\endcsname\xspace}%
  \expandafter\newcommand\csname #1\endcsname[1]%
  {\ensuremath{\csname #1sym\endcsname #3{##1}}\xspace}%
}

\newcommand{\encloseinpar}[1]{(#1)}

\newcommand{\encloseinsembra}[1]{|[#1|]}
\newcommand{\encloseinangles}[1]{\langle #1\rangle}

\newcommand{\mksubscript}[1]{_{#1}}
\newcommand{\donotenclose}[1]{#1}
\newcommand{\followwithbang}[1]{#1{\ensuremath{\scriptstyle !}}}
\newcommand{\followwithqstn}[1]{#1{\ensuremath{\scriptstyle ?}}}
\newcommand{\followwithnot}[1]{#1{\raisebox{-0.8ex}[0mm][.1ex]{\ensuremath{\scriptstyle \!\not\,\,}}}}
\newcommand{\followwithast}[1]{#1^\ast}

\newcommand*{\mkfuncn}[2]{\mkfunc{#1}{#2}{\donotenclose}}    
\newcommand*{\mkfuncp}[2]{\mkfunc{#1}{#2}{\encloseinpar}}    
\newcommand*{\mkfuncs}[2]{\mkfunc{#1}{#2}{\encloseinsembra}} 
\newcommand*{\mkfuncsub}[2]{\mkfunc{#1}{#2}{\mksubscript}}   

\newcommand*{\mkbinaryrel}[2]{%
  \expandafter\newcommand\csname#1sym\endcsname%
  {{\ensuremath{#2}}}%
  \expandafter\newcommand\csname not#1sym\endcsname%
  {{\ensuremath{\not{#2}}}}%
  \expandafter\newcommand\csname #1name\endcsname%
  {\csname #1sym\endcsname\xspace}%
  \expandafter\newcommand\csname not#1name\endcsname%
  {\csname not#1sym\endcsname\xspace}%
  \expandafter\newcommand\csname #1\endcsname[2]%
  {{\ensuremath{##1#2##2}}}%
  \expandafter\newcommand\csname not#1\endcsname[2]%
  {{\ensuremath{##1 \not#2  ##2}}}%
}

\newcommand*{\mkternaryrel}[2]{%
  \expandafter\newcommand\csname #1sym\endcsname%
  {\ensuremath{#2}}%
  \expandafter\newcommand\csname not#1sym\endcsname%
  {\ensuremath{\not #2}}%
  \expandafter\newcommand\csname #1name\endcsname%
  {\csname #1sym\endcsname\xspace}%
  \expandafter\newcommand\csname not#1name\endcsname%
  {\csname not#1sym\endcsname\xspace}%
  \expandafter\newcommand\csname #1\endcsname[3]%
  {\ensuremath{##1 \csname #1sym\endcsname_{##2} ##3}}%
  \expandafter\newcommand\csname not#1\endcsname[3]%
  {\ensuremath{##1 \not\csname #1sym\endcsname_{##2} ##3}}%
}

\newcommand*{\mkternaryrelraised}[2]{%
  \expandafter\newcommand\csname #1sym\endcsname%
  {\ensuremath{#2}}%
  \expandafter\newcommand\csname not#1sym\endcsname%
  {\ensuremath{\not #2}}%
  \expandafter\newcommand\csname #1name\endcsname%
  {\csname #1sym\endcsname\xspace}%
  \expandafter\newcommand\csname not#1name\endcsname%
  {\csname not#1sym\endcsname\xspace}%
  \expandafter\newcommand\csname #1\endcsname[3]%
  {\ensuremath{##1 \csname #1sym\endcsname{}^{##2} ##3}}%
  \expandafter\newcommand\csname not#1\endcsname[3]%
  {\ensuremath{##1 \not\csname #1sym\endcsname{}^{##2} ##3}}%
}

\newcommand*{\mkcomrel}[5]{%
  \ifgiven{#4}\then%
    \expandafter\newcommand\csname #1sym\endcsname%
    {\raisebox{.2ex}{\ensuremath{\xrightarrow[{\raisebox{.5ex}[0mm][.1ex]{\tiny #4}}]%
          {#3{}}}}{}%
    \index{#4@\protect{\csname #1sym\endcsname}}}%
    \expandafter\newcommand\csname #1name\endcsname%
    {\csname #1sym\endcsname\xspace}%
    \expandafter\newcommand\csname #1\endcsname[3]%
    {\hbox{\ensuremath{%
          #2{##1}%
          \raisebox{-.1ex}{\ensuremath{\xrightarrow[{\raisebox{.5ex}[0mm][.1ex]{\smash{\tiny #4}}}]%
              {\phantom{.}\raisebox{-0.3ex}[0mm][0.1ex]{{#3{\ensuremath{\scriptstyle ##2}}}}\phantom{.}}}}%
          {}#5{##3}%
        }%
      }}%
  \else%
    \expandafter\newcommand\csname #1sym\endcsname%
    {\ensuremath{\xrightarrow{#3{}}}%
      \index{#4@\protect{\csname #1sym\endcsname}}}%
    \expandafter\newcommand\csname #1name\endcsname%
    {\csname #1sym\endcsname\xspace}%
    \expandafter\newcommand\csname #1\endcsname[3]%
    {\hbox{\ensuremath{%
          #2{##1}%
          \ensuremath{\xrightarrow{#3{##2}}}%
          #5{##3}%
        }%
      }\index{#4@\protect{\csname #1sym\endcsname}}}%
  \fi%
}

\newcommand*{\mkoperel}[4]{%
  \expandafter\newcommand\csname #1sym\endcsname%
       {\raisebox{.2ex}{\ensuremath{\xrightarrow[{\raisebox{.5ex}[0mm][.1ex]{\tiny #3}}]%
             {}}}\index{#3@\protect{\csname #1sym\endcsname}}}%
  \expandafter\newcommand\csname #1name\endcsname%
       {\csname #1sym\endcsname\xspace}%
  \expandafter\newcommand\csname #1\endcsname[2]%
       {\hbox{\ensuremath{%
              #2{##1}%
              \raisebox{.2ex}{\ensuremath{\xrightarrow[{\raisebox{.5ex}[0mm][.1ex]{\tiny #3}}]{}}}%
              #4{##2}%
            }%
          }\index{#3@\protect{\csname #1sym\endcsname}}}%
}

\newcommand*{\mkoperelaa}[2]{%
  \mkoperel{#1}{\encloseinangles}{#2}{\encloseinangles}}                      
\newcommand*{\mkcomrelaa}[2]{%
  \mkcomrel{#1}{\encloseinangles}{\donotenclose}{#2}{\encloseinangles}}   
\newcommand*{\mkoperelnn}[2]{%
  \mkoperel{#1}{\donotenclose}{#2}{\donotenclose}}                            
\newcommand*{\mkcomrelnn}[2]{%
  \mkcomrel{#1}{\donotenclose}{\donotenclose}{#2}{\donotenclose}}         
\newcommand*{\mkoperelan}[2]{%
  \mkoperel{#1}{\encloseinangles}{#2}{\donotenclose}}                         
\newcommand*{\mkcomrelan}[2]{%
  \mkcomrel{#1}{\encloseinangles}{\donotenclose}{#2}{\donotenclose}}      

\newcommand*{\mkdirrelaa}[3]{%
  \mkcomrel{#1}{\donotenclose}{#2}{#3}{\donotenclose}}         

\newcommand{\VSsym}{\index{visualSTATE@\protect{\VSsym}}\textsf{visual\-STATE}}

\newcommand{\STM}{\index{Statemate@\protect{\STM}}\textsc{statemate}\xspace}
\newcommand{\SCsym}{\expandafter\index{SCOPE@\protect{\SCsym}}\textsf{SCOPE}}

\newcommand{\Rhapsodysym}{\index{Rhapsody@\protect{\Rhapsodysym}}\textsc{Rhapsody}}

\newcommand{\inonek}{\hbox{\ensuremath{\in\kern-.5mm\{1..k\}}}}

\mkfuncp{elems}{\wordsp{elems}}
\mkfuncp{dom}{\wordsp{dom}}
\mkfuncp{rng}{\wordsp{rng}}
\mkfuncp{Power}{\mathscr{P}}
\mkfuncp{Multi}{\mathcal{M}}
\mkfuncp{Forest}{\mathcal{F}}

\mkfunc{List}{}{\followwithast}
\newcommand{\Projectsym}[1]{\ensuremath{\pi_{#1}}}
\newcommand{\Project}[2]
        {\ensuremath{\Projectsym{#1}({#2})}\xspace}

\mktype{OR}{or}             
\mktype{XOR}{xor}           
\mktype{AND}{and}    

\mkkeyword{Ids}{\wordsp{Id}}
\mkkeyword{State}{\wordsp{State}}
\mkkeyword{History}{\wordsp{History}}

\newcommand*{\ORst}{\expandafter\index{or-state@\protect{\ORst}}\ORsym{}-state\xspace}
\newcommand*{\ORsts}{\expandafter\index{or-state@\protect{\ORst}}\ORsym{}-states\xspace}
\newcommand*{\XORst}{\expandafter\index{xor-state@\protect{\XORst}}\XORsym{}-state\xspace}
\newcommand*{\XORsts}{\expandafter\index{xor-state@\protect{\XORst}}\XORsym{}-states\xspace}
\newcommand*{\ANDst}{\expandafter\index{and-state@\protect{\ANDst}}\ANDsym{}-state\xspace}
\newcommand*{\ANDsts}{\expandafter\index{and-state@\protect{\ANDst}}\ANDsym{}-states\xspace}

\newcommand*{\StateORsym}{\index{\protect{\StateORsym}}\ensuremath{\State_{\text{\ORsym}}}}   
\newcommand*{\StateANDsym}{\index{\protect{\StateANDsym}}\ensuremath{\State_{\text{\ANDsym}}}}

\mkfuncp{type}{\Gamma_S}

\newcommand{\sbstatekern}{\kern -0.18em}

\newcommand{\sbstatesym}{\ensuremath{\searrow}\xspace}  
\newcommand{\notsbstatesym}{\ensuremath{\not\searrow}\xspace}

\newcommand{\sbstateplsym}{\ensuremath{\sbstatesym\kern -0.66em^{+}\kern 0.13em}}
 
\newcommand{\notsbstatepl}[2]{\hbox{\ensuremath{#2\sbstatekern\notsbstatesym\kern -0.66em^{+}#1}}\xspace} 
\newcommand{\sbstatestsym}{\ensuremath{\sbstatesym\kern -0.66em^\ast\kern 0.13em}}
 
\newcommand{\notsbstatest}[2]{\hbox{\ensuremath{#2\sbstatekern\notsbstatesym\kern -0.66em^\ast#1}}\xspace}

\mkfuncp{priority}{\vartriangleleft}
\mkfuncp{stpriority}{\blacktriangleleft}

\mkfuncp{parent}{\wordsp{parent}}
\mkfuncp{parparent}{\wordsp{parent}^2}
\mkfuncp{children}{\wordsp{children}}
\mkfuncp{ancestors}{\wordsp{ancest}}
\mkfuncp{descendants}{\wordsp{descend}}
\mkfuncp{siblings}{\wordsp{siblings}}

\mkfuncp{ancestorspl}{\wordsp\ancestorssym^{+}}
\mkfuncp{descendantspl}{\wordsp\descendantssym^{+}}
\mkfuncp{ancestorsst}{\wordsp\ancestorssym^\ast}
\mkfuncp{descendantsst}{\wordsp\descendantssym^\ast}
\mkfuncp{siblingsst}{\wordsp{\siblingssym}^\ast}

\mkfuncp{NCA}{\wordsp{NCA}}

\newcommand{\notorthogonalsym}{\hbox{\ensuremath{\not{\kern -2.2pt\bot\kern +2.2pt}}}}

\mkfuncp{initial}{\wordsp{ini}}
\mkfuncp{default}{\wordsp{default}}
\mkfuncp{runhistory}{\eta}

\mkfuncp{entrypathOR}{\entrypathsym_{\text{\ORsym}}}
\mkfuncp{entrypathAND}{\entrypathsym_{\text{\ANDsym}}}
\mkfuncp{entrypath}{\wordsp{entr}}
\mkfuncp{entrysigs}{\wordsp{ensigs}}

\mkfuncp{entrORts}{\entrypathORsym\kern-6pt^{\wordsp{ts}}}
\mkfuncp{entrANDts}{\entrypathANDsym\kern-10pt^{\wordsp{ts~}}}
\mkfuncp{exitpath}{\wordsp{extr}}
\mkfuncp{exitsigs}{\wordsp{exsigs}}

\mkkeyword{Id}{\wordsp{Id}}
\mkkeyword{Var}{\wordsp{Var}}
\mkkeyword{Value}{\wordsp{Value}}

\mkkeyword{ExtVar}{\Var_E}
\mkkeyword{IntVar}{\Var_I}
\mkkeyword{Type}{\wordsp{Type}}
\mkkeyword{SimpleType}{\wordsp{SimpleType}}

\mkkeyword{anatomy}{\wordsp{anatomy}}

\mktype{tint}{int}
\mktype{tuint}{uint}
\mktype{treal}{real}
\mktype{tfloat}{float}
\mktype{tdouble}{double}
\mktype{tvoid}{void}

\mkfuncp{Dom}{D}

\mkfuncp{vartype}{\Gamma_V}
\mkfuncp{Array}{\wordsp{Vector}}
\mkfuncs{toracle}{\tau}

\mkfuncp{mustclosure}{\raisebox{-2pt}{$\Box$}\textit{-closure}}

\mkkeyword{Event}{\wordsp{Event}}
\mkkeyword{Signal}{\wordsp{Signal}} 
\mkkeyword{External}{\wordsp{External}} 
\mkkeyword{Guard}{\wordsp{Guard}}
\mkkeyword{Ebind}{\wordsp{Ebind}}
\mkkeyword{Einst}{\wordsp{Einst}}
\mkkeyword{Sexp}{\wordsp{Sexpr}}
\mkkeyword{Lvalue}{\wordsp{Access}}

\mkfuncp{evtype}{\Gamma_E}

\mkkeyword{topstate}{\wordsp{root}}

\mkkeyword{Exp}{\wordsp{Exp}}
\mkkeyword{OExp}{\wordsp{exp}}
\mkkeyword{Aexp}{\wordsp{Aexp}}
\mkkeyword{Bexp}{\wordsp{Bexp}}

\mkkeyword{Assgn}{\wordsp{Assgn}}
\mkkeyword{Output}{\wordsp{Output}}
\mkkeyword{Oinst}{\wordsp{Oinst}}
\mkkeyword{Ainst}{\wordsp{Ainst}}
\mkkeyword{Oexp}{\wordsp{Oexpr}}

\mkfuncp{fntype}{\Gamma_F}

\mkkeyword{Action}{\wordsp{Action}}
\mkkeyword{Trans}{\wordsp{Trans}}

\mkfuncp{event}{\wordsp{event}}
\mkfuncp{params}{\wordsp{params}}
\mkfuncp{scope}{\wordsp{scope}}
\mkfuncp{iscope}{\wordsp{iscope}}
\mkfuncp{scopes}{\wordsp{scope}}
\mkfuncp{poscond}{\wordsp{pos}}
\mkfuncp{negcond}{\wordsp{neg}}
\mkfuncp{targets}{\wordsp{targets}}
\mkfuncp{source}{\wordsp{source}}
\mkfuncp{expr}{\wordsp{expr}}
\mkfuncp{guard}{\wordsp{guard}}
\mkfuncp{entry}{\wordsp{en}}
\mkfuncp{exit}{\wordsp{ex}}

\mkfuncp{freevars}{\wordsp{fvars}}
\mkfuncp{pure}{\wordsp{pure}}
\mkfuncp{closed}{\wordsp{closed}}

\mkkeyword{Store}{\wordsp{Store}}
\mkkeyword{Queue}{\wordsp{Queue}}
\mkkeyword{state}{\sigma}
\mkkeyword{gstate}{\Sigma}

\mkfuncp{store}{\varrho}
\mkfuncp{history}{\wordsp{his}}

\mkkeyword{queue}{q}

\mkoperelnn{trarrow}{\stignore{0transitionrelation}}
\mkoperelaa{access}{\textrm{access}}
\mkoperelaa{Aeval}{\textrm{Aeval}}
\mkoperelan{Aexec}{\textrm{asgn}}
\mkoperelan{Beval}{\textrm{Beval}}
\mkoperelaa{Cexec}{\textrm{C-sem}}
\mkcomrelnn{comarrow}{}
\mkcomrelnn{Init}{\textrm{init}}
\mkcomrelaa{Macro}{\textrm{macro}}
\mkcomrelaa{Micro}{\textrm{micro}}
\mkoperelaa{Mini}{\textrm{mini}}
\mkcomrelaa{Oeval}{\textrm{output}}
\mkcomrelan{Rexec}{\textrm{exec}}  
\mkcomrelnn{Senter}{\textrm{enter}}
\mkoperelaa{Seval}{\textrm{signal}}
\mkcomrelnn{Sexit}{\textrm{exit}}
\mkcomrelaa{systr}{\stignore{0transitionrelation}}
\mkcomrelnn{Tfire}{\textrm{fire}}

\mkfuncp{cons}{\wordsp{cons}}

\mkfuncn{tail}{\text{\bfseries tl}\,}

\mkfuncn{etail}{{\textbf{src}}\,}
\mkfuncn{ehead}{{\textbf{tgt}}\,}

\mkfuncp{Br}{\textsf{Br}}

\mkfuncp{activeset}{\wordsp{enabled}}
\mkfuncp{enabled}{\wordsp{enabled}}
\mkfuncp{resactive}{\wordsp{resolved-enabled}}
\mkfuncp{activestate}{\wordsp{active}}
\mkfuncp{inactivestate}{\wordsp{inactive}}

\mkfuncp{traces}{\wordsp{traces}}

\mkkeyword{PR}{\wordsp{Pr}}
\mkkeyword{PRP}{\PRsym_{\mathcal{P}}}
\mkkeyword{PRS}{\PRsym_{\ast}}
\mkkeyword{ENV}{\wordsp{Env}}
\mkkeyword{GEN}{\wordsp{Gen}}
\mkkeyword{GENP}{\GENsym_\mathcal{P}}
\mkkeyword{GENS}{\GENsym_\ast}
\mkkeyword{OBS}{\wordsp{Obs}}
\mkkeyword{OBSP}{\OBSsym_\mathcal{P}}
\mkkeyword{OBSS}{{\OBSsym_\ast}}

\mkkeyword{OUT}{\wordsp{Out}}
\mkkeyword{IN}{\wordsp{In}}

\newcommand{\gentrsymname}{}
\newcommand{\obstrsymname}{}
\newcommand{\nottrsymname}{}
\mkdirrelaa{gentr}{\followwithbang}{\gentrsymname}
\mkdirrelaa{obstr}{\followwithqstn}{\obstrsymname}
\mkdirrelaa{nottr}{\followwithnot}{\nottrsymname}

\mkternaryrel{relsim}{\leqslant}
\mkternaryrel{relbisim}{\sim}
\mkternaryrel{twowayrelsim}{\index{simulation!relativized, two way}\lessgtr}
\mkbinaryrel{twowaysim}{\index{simulation!two way}\lessgtr}
\mkbinaryrel{osim}{\leqslant}
\mkbinaryrel{gsim}{\leqslant}
\mkbinaryrel{obisim}{\sim}
\mkbinaryrel{gbisim}{\sim}
\mkbinaryrel{syssim}{\leqslant}
\mkbinaryrel{modref}{\leq_{\textnormal{\textrm{m}}}}
\mkbinaryrel{altsim}{\leq_{\textnormal{\textrm{a}}}}
\mkternaryrelraised{omodref}{\leq^{\ast}_{\textnormal{\textrm{m}}}}
\mkternaryrelraised{agmodref}{\leq^{\ast}_{\textnormal{\textrm{ag}}}}
\mkternaryrelraised{esopref}{\leq^{\triangleleft}_{\textnormal{\textrm{m}}}}
\mkbinaryrel{oaltsim}{\leq^{\ast}_{\textnormal{\textrm{a}}}}
\mkbinaryrel{envsim}{\leqslant}
\mkbinaryrel{sysbisim}{\sim}
\mkbinaryrel{discr}{\index{discrimination}\sqsubseteq}
\mkbinaryrel{implements}{\preceq}
\mkbinaryrel{twodiscr}{\raisebox{1.25ex}{\rotatebox[origin=ct]{-180}%
    {\index{discrimination}\smash{\ensuremath{\sqsupseteq}}}}}
\mkbinaryrel{bidiscr}{\raisebox{-.5ex}{\ensuremath{%
      \index{discrimination}\stackrel{\sqsubset}{\scalebox{0.8}{\ensuremath{\sim}}}}}}

\mkbinaryrel{composetr}{\!\upharpoonright\!}
\mkbinaryrel{crossprod}{\otimes}

\mkkeyword{IOATS}{\text{\index{IOATS}IOATS}}
\mkkeyword{IOATSs}{\text{\index{IOATS}IOATSs}}

\mkfuncp{SSF}{\mathbb{S}}

\mkkeyword{SET}{\textbf{Set}}
\mkkeyword{set}{\textbf{set}}
\mkkeyword{MSET}{\textbf{M-set}}
\mkkeyword{mset}{\textbf{m-set}}
\mkkeyword{INTER}{\textnormal{\textbf{Inter}}}
\mkkeyword{inter}{\textnormal{\textbf{inter}}}

\newcommand{\BLINDsym}{\ensuremath{\index{environment!blind}\mathbf{B}}}

\mkfuncsub{parBLIND}{\BLINDsym}

\newcommand{\UNIVsym}{{\ensuremath{\index{environment!perfect vision}\mathbf{V}}}}

\mkfuncsub{parUNIV}{\UNIVsym}

\mkfuncp{eqdom}{\wordsp{part}}

\usepackage{psfrag}
\usepackage{ifthen}
\usepackage[british]{babel}

\usepackage{xspace}  
\newcommand{\prism}{\textsc{Prism}\xspace}
 
\newcommand{\WBS}[2][\null]
    {\ensuremath{\textit{WBS}\left(#2,
     \ifthenelse{\equal{#1}{\null}}{\emptyset}{#1}\right)}\xspace}

\newcommand{\St}{\textit{St}\xspace}

\newcommand{\xRightarrow}[2][]{%
     \ext@arrow 0055{\Rightarrowfill@}{#1}{#2}%
} 

\renewcommand{\St}{{\sf St}}
\newcommand{\A}{\boldsymbol{{\cal A}}}

\renewcommand{\phi}{\varphi}

\newcommand{\uppaal}{\textsc{Uppaal}\xspace}

\newcommand{\forget}[1]{}

\newcommand{\bbbn}{\ensuremath{{\mathbb N}}\xspace}
\newcommand{\bbbr}{\ensuremath{{\mathbb R}}\xspace}

\newcommand{\rplus}{\ensuremath{\bbbr_{\geq0}}}

\newcommand{\U}{{\cal U}}
\renewcommand{\L}{{\cal L}}
\newcommand{\arrow}[1]{\stackrel{#1}{\longrightarrow}}

\theoremstyle{plain}{\theorembodyfont{\rmfamily}\newtheorem{definition}{Definition}}
\theoremstyle{plain}{\theorembodyfont{\rmfamily}}
\theoremstyle{plain}{\theorembodyfont{\rmfamily}\newtheorem{theorem}{Theorem}}
\theoremstyle{plain}{\theorembodyfont{\rmfamily}}
\theoremstyle{plain}{\theorembodyfont{\rmfamily}}
\theoremstyle{plain}{\theorembodyfont{\rmfamily}}
\theoremstyle{plain}{\theorembodyfont{\rmfamily}\newtheorem{example}{Example}}

\hypersetup{
  pdftitle={Stochastic Semantics and Statistical Model Checking for Networks of Priced Timed Automata},
  pdfsubject={},
  pdfkeywords={timed automata, hybrid systems, model checking, statistical model checking, simulation},
  pdfauthor={Alexandre David, Kim. G. Larsen, Axel Legay, Marius Mikucionis, Danny Boegsted Poulsen, Jonas van Vliet, Zheng Wang},
  linkcolor={black!50!red},
  citecolor={black!50!green},
  urlcolor={black!50!blue}
}

\begin{document}

\title{
Stochastic Semantics and Statistical Model Checking for Networks of Priced Timed Automata
\thanks{
Work partially supported by VKR Centre of Excellence -- MT-LAB
and by an ``Action de Recherche Collaborative'' ARC (TP)I.}
}

\author{
\IEEEauthorblockN{Alexandre~David, Kim~G.~Larsen,\\
    Marius~Mikučionis, Danny~Bøgsted~Poulsen,\\
 and Jonas~van~Vliet}
\IEEEauthorblockA{Department of Computer Science\\
Aalborg University, Denmark\\
Email: \{adavid,kgl,marius,dannypb,jonasvv\}@cs.aau.dk}
\and
\IEEEauthorblockN{Axel~Legay}
\IEEEauthorblockA{INRIA/IRISA\\Rennes Cedex, France\\
Email: axel.legay@irisa.fr}
\and
\IEEEauthorblockN{Zheng~Wang}
\IEEEauthorblockA{Software Engineering Institute\\
East China Normal University, China\\
Email: wangzheng@sei.ecnu.edu.cn}
}

\maketitle

\begin{abstract}
  This  paper offers  a natural  stochastic semantics  of  Networks of
  Priced Timed Automata (NPTA)  based on races between components. The
  semantics  provides  the  basis  for satisfaction  of  Probabilistic
  Weighted  CTL  properties   (PWCTL),  conservatively  extending  the
  classical satisfaction  of timed automata with respect  to TCTL.  In
  particular  the extension  allows for  hard real-time  properties of
  timed  automata expressible  in TCTL  to be  refined  by performance
  properties, e.g.  in terms  of probabilistic guarantees of time- and
  cost-bounded properties.  A second  contribution of the paper is the
  application  of  Statistical  Model  Checking (SMC)  to  efficiently
  estimate the correctness of non-nested PWCTL model checking problems
  with a desired level of confidence, based on a number of independent
  runs of the NPTA.  In addition to applying classical SMC algorithms,
  we  also  offer an  extension  that  allows  to efficiently  compare
  performance properties of NPTAs  in a parametric setting.  The third
  contribution is  an efficient tool implementation of  our result and
  applications to several case studies.
\end{abstract}

\IEEEpeerreviewmaketitle

\section{Introduction}
\fxnote{Axel}

Model Checking  (MC)\,\cite{CGP99} is a widely  recognised approach to
guarantee  the correctness of  a system  by checking  that any  of its
behaviors is a model for a given property.  There are several variants
and extensions of  MC aiming at handling real-time  and hybrid systems
with  quantitative  constraints  on   time,  energy  or  more  general
continuous aspects \cite{Alur-Dill-94,ACHH95,PTA01,WTA01}.  Within the
field of embedded systems  these formalisms and their supporting tools
\cite{SPIN,SMV,UPPAAL,Phaver} are now successfully applied to time- and
energy-optimal scheduling, WCET analysis and schedulability analysis.

Compared  with traditional  approaches,  a strong  point of  real-time
model checking is  that it (in principle) only requires  a model to be
applicable, thus extensions to multi-processor setting is easy. A
weak point of  model checking is the notorious  problem of state-space
explosion, i.e. the exponential growth in the analysis effort measured
in the  number of  model-components.  Another limitation  of real-time
model checking is that it merely provides -- admittedly most important
-- hard quantitative guarantees, e.g.  the worst case response time of
a recurrent task under a  certain scheduling principle, the worst case
execution time  of a  piece of code  running on a  particular execution
platform,  or the worst  case time  before consensus  is reached  by a
real-time network  protocol.  In addition to these  hard guarantees, it
would be desirable in  several situations to obtain  refined performance
information concerning likely or expected behaviors in terms of timing
and  resource   consumption.  In   particular,  this  would   allow  to
distinguish and select between systems that perform identically from a
worst-case perspective.

\begin{figure*}[t]
\begin{center}
\begin{tabular}{ll}
a) Axel & \includegraphics[height=0.11\textheight]{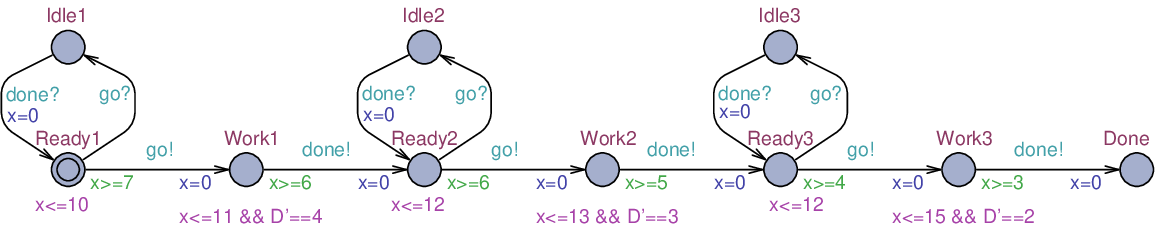}\\
b) Alex &\includegraphics[height=0.11\textheight]{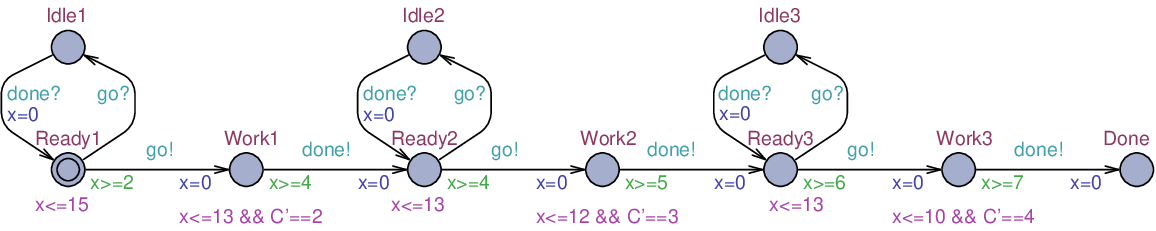} 
\end{tabular}
\end{center}
\caption{3-Nail Hammering Game between Axel and  Alex.}
\label{fig:hammer}
\end{figure*}

\begin{figure*}[t] 
a)\includegraphics[width=0.48\textwidth]{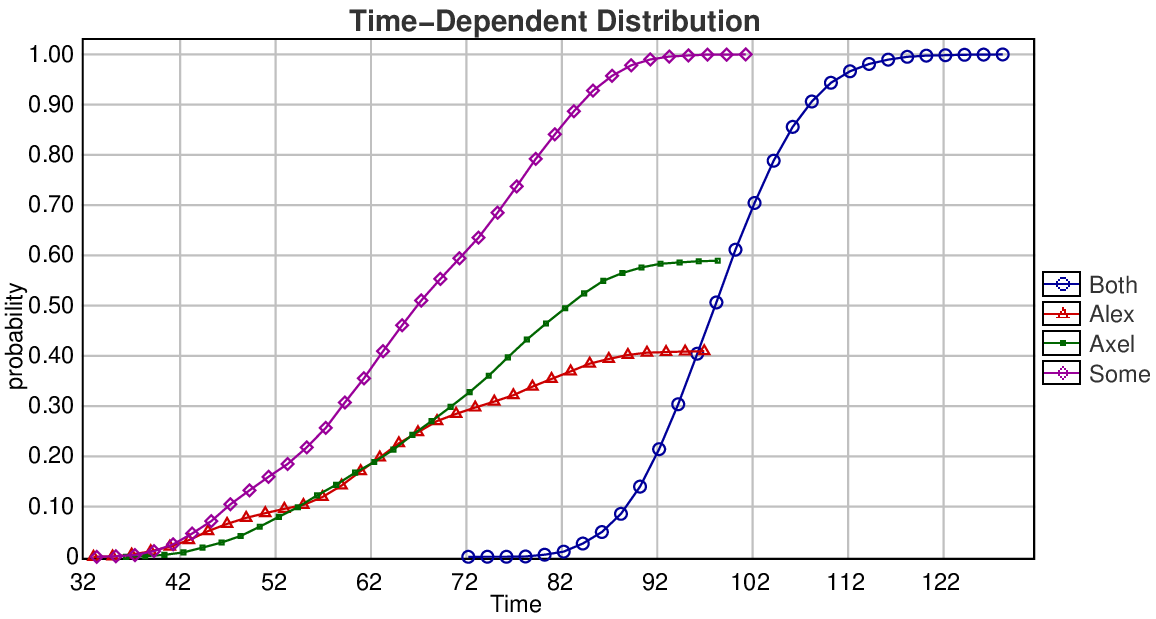} \hspace{\stretch{1}}
b)\includegraphics[width=0.48\textwidth]{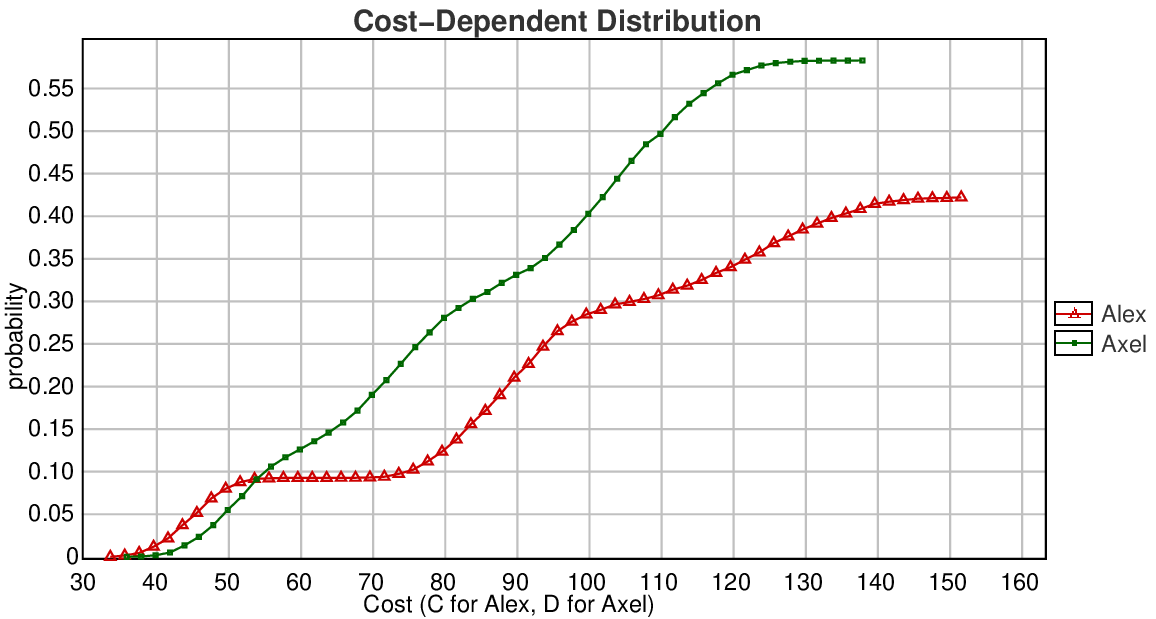}
\caption{Time- and Cost-dependent Probability of winning the
  Hammering Game}
\label{fig:TimeCost}
\end{figure*}

To  illustrate our  point consider  the  network of  two priced  timed
automata in Fig.~\ref{fig:hammer} modeling a competition between Axel
and Alex  both having to hammer three  nails down.  As can  be seen by
the representing {\tt Work}-locations the time (-interval) and rate of
energy-consumption required for hammering a nail depends on the player
and the  nail-number.  As  expected Axel is  initially quite  fast and
uses a lot of energy but  becomes slow towards the last nail, somewhat
in contrast to  Alex. To make it an  interesting competition, there is
only  \emph{one} hammer illustrated  by repeated  competitions between
the two players in the {\tt Ready}-locations, where the slowest player
has to  wait in  the {\tt Idle}-location  until the faster  player has
finished hammering the next  nail. Interestingly, despite the somewhat
different strategy applied, the  best- and worst-case completion times
are identical for Axel and Alex: 59 seconds and 150 seconds. So, there
is  no difference between  the two  players and  their strategy,  or
is there?
Assume that  a third  person wants to  bet on  who is the  more likely
winner  --  Axel or  Alex  -- given  a  refined  semantics, where  the
time-delay before performing an output is chosen stochastically (e.g.
by  drawing  from  a  uniform  distribution).  Under  such  a  refined
semantics there  is a significant difference between  the two players.
In Fig.~\ref{fig:TimeCost}a) the  probability distributions for either
of the two players winning before  a certain time is given.  Though it
is clear that Axel has a higher probability of winning than Alex (59\%
versus 41\%), however declaring the competition a draw if it has not finished
before  50  seconds  actually  makes  Alex  the  more  likely  winner.
Similarly,  Fig.~\ref{fig:TimeCost}b)  illustrates  the probability  of
either  of the two  players winning  given an  upper bound  on energy.
With an  unlimited amount of energy,  clearly Axel is  the most likely
winner,  whereas limiting  the  consumption of  energy  to maximum  52
``energy-units'' gives Alex an advantage.

As a first contribution of this paper we propose a stochastic
semantics for Priced Timed Automata (PTA), whose clocks can evolve
with different rates, while\footnote{in contrast to the usual
  restriction of priced timed automata \cite{PTA01,WTA01}} being used
with no restrictions in guards and invariants.  Networks of PTAs
(NPTA) are created by composing PTAs via input and output actions.
The model is as expressive as linear hybrid automata\,\cite{ACHH95},
making even the reachability problem undecidable. More precisely, we
define a natural stochastic semantics for networks of NPTAs based on
races between components. We shall observe that such race can generate
arbitrarily complex stochastic behaviors from simple assumptions on
individual components. While fully stochastic semantics have already
been proposed for timed systems \cite{BBBBG07,BBBM08}, we are the
first to consider networks of timed and hybrid systems.  Other related
work includes the very rich framework of stochastic timed systems of
MoDeST \cite{BDArHK04}.  Here, however, general hybrid variables are
not considered and parallel composition does not yield fully
stochastic models.  For the notion of probabilistic hybrid systems
considered in \cite{HSCC11} the choice of time is resolved
non-deterministically rather than stochastically as in our case.
Moreover, based on the stochastic semantics, we are able to express
refined performance properties, e.g.  in terms of probabilistic
guarantees of time- and cost-bounded properties.

To  allow  for the  efficient  analysis  of probabilistic  performance
properties  --  despite the  general  undecidability  of  these --  we
propose     to     work     with    Statistical     Model     Checking
(SMC)\,\cite{You05a,SVA04},  an   approach  that  has   recently  been
proposed as an  alternative to avoid an exhaustive  exploration of the
state-space of the model. The core  idea of the approach is to monitor
some  simulations  of  the  system,  and then  use  results  from  the
statistic area (including sequential hypothesis testing or Monte Carlo
simulation)  in  order to  decide  whether  the  system satisfies  the
property or not  with some degree of confidence.  By  nature, SMC is a
compromise between testing and classical model checking techniques.


Thus, as a second contribution, we provide an efficient implementation
of existing SMC algorithms that we use for checking the correctness of
NPTAs with respect to cost-constrained temporal logic. The series of
algorithms we implement includes a version of the sequential
hypothesis test by Wald\,\cite{Wal04} as well as a quantitative
approach\,\cite{HLMP04}. Our implementation relies on a new efficient
algorithm for generating runs of NPTAs in a random manner.  In
addition, we also propose another SMC algorithm to compare the
performances of two properties without computing their
probability. This problem, which is far beyond the scope of existing
time model checking approaches, can be approximated with an extension
of the sequential hypothesis testing. In addition to be the first to
apply such extension in the context of formal verification, we also
propose a new variant that allows to reuse existing results in
parallel when comparing the properties on different timed bounds.

Finally, one of the most interesting contribution of our work takes
the form of a series of new case studies that are analyzed with a new
stochastic extension of \uppaal\,\cite{DLLMW11}. Particularly, we show
how our approach can be used to resolve scheduling problems. Such
problems are defined using Duration Probabilistic Automata
(DPA)\,\cite{MLK10}, a new and natural model for specifying list of
tasks and shared resources. We observe that our approach is not only
more general, but also much faster than the hypothesis testing engine
recently implemented in the \prism toolset. Our work thus presents
significant advances in both the modeling and the efficient
verification of network of complex systems.

{\bf Related work.} Some works on probabilistic semantics of timed 
automata have already been discussed above. Simulation-based
approaches such as Monte Carlo have been in use since decades, however 
the use of simulation and hypothesis testing to reason on formal models 
is a more recent advance. First attempts to apply hypothesis testing 
on stochastic extension of Hennessy-Milner logic can be found in~\cite{LS89}.
In~\cite{YS02,You05a}, Younes was the first to apply hypothesis testing
to stochastic systems whose properties are specified with (bounded)
temporal logic. His approach is implemented in the Ymer
toolset~\cite{You05c} and can be applied on time-homogeneous generalized 
semi-Markov processes, while our semantics addresses the composition of
stochastic systems allowing to compose a global system from components
and reason about communication between independent processes. In
addition to Younes work we explore continuous-time features, 
formalize and implement Wald's ideas where the probability comparison 
can be evaluated on NPTA processes. In a recent work~\cite{ZPC10}, 
Zuliani et al. extended the SMC approach to hybrid systems. 
Their work is a combination of~\cite{JCLLPZ09} and~\cite{CDL08} based 
on Simulink models (non-linear hybrid systems), whereas our method 
is specialised to networks of priced timed automata where model-checking 
techniques can be directly applicable using the same tool suite. 
In addition we provide means of comparing performances without 
considering individual probabilities. 
Finally, a very recent work~\cite{BFHH11} proposes partial order reduction 
techniques to resolve non-determinism between components rather than 
defining a unique stochastic distribution on their product behaviors.
While this work is of clear interest, we point out that the
application of partial order may considerably increase the computation
time and for some models partial orders cannot resolve non-determinism, 
especially when considering continuous time~\cite{Min99}. 
Other works on SMC can be found in~\cite{SVA05,BBBCDL10}.

\section{Network of Priced Timed Automata}
\label{sec:CSTA}
\fxnote{Kim}

We consider the notion of {\em Networks of Priced Timed Automata
  (NPTA)}, generalizing that of regular timed automata (TA) in that
clocks may have different rates in different locations.
In fact, the expressive power (up to timed bisimilarity) of NPTA
equals that of general linear hybrid automata (LHA) \cite{ACHH95},
rendering most problems -- including that of reachability --
undecidable.


Let $X$ be a finite set of variables, called \emph{clocks}\footnote{We
  will (mis)use the term ``clock''  from timed automata, though in the
  setting of NPTAs the variables in $X$ are really general real-valued
  variables.}.   A  \emph{clock  valuation}  over  $X$  is  a  mapping
$\nu:X\rightarrow \rplus$,  where $\rplus$  is the set  of nonnegative
reals.  We write $\rplus^X$ for the set of clock valuations over
$X$.  Let $r:X\rightarrow\bbbn$ be  a \emph{rate vector}, assigning to
each clock of $X$ a rate.  Then, for $\nu\in\rplus^X$ and $d\in\rplus$
a delay,  we write $\nu+r\cdot d$  for the clock  valuation defined by
$(\nu+ r\cdot  d)(x)=\nu(x)+r(x)\cdot d$ for  any clock $x\in  X$.  We
denote by $\bbbn^X$  the set of all rate  vectors.  If $Y\subseteq X$,
the valuation  $\nu[Y]$ is the  valuation assigning $0$ when  $x\in Y$
and $\nu(x)$ when $x\not\in  Y$.  An \emph{upper bounded (lower bound)
  guard} over  $X$ is a finite  conjunction of simple  clock bounds of
the   form    $x\sim   n$    where   $x\in   X$,    $n\in\bbbn$,   and
$\sim\in\{<,\leq\}$   ($\sim\in\{>,\geq\}$)  We   denote   by  $\U(X)$
($\L(X)$ the  set of  upper (lower) bound  guards over $X$,  and write
$\nu\models  g$ whenever  $\nu$ is  a clock  valuation  satisfying the
guard $g$.  Let $\Sigma=\Sigma_i\uplus\Sigma_o$  be a disjoint sets of
input and output actions.

\begin{definition} A \emph{Priced Timed Automaton} (PTA) is a
  tuple  ${\cal  A}=(L,\ell_0,X,\Sigma,E,R,I)$  where:  (i) $L$  is  a
  finite set of locations, (ii) $\ell_0\in L$ is the initial location,
  (iii)    $X$     is    a     finite    set    of     clocks,    (iv)
  $\Sigma=\Sigma_i\uplus\Sigma_o$   is  a   finite   set  of   actions
  partitioned into  inputs ($\Sigma_i$) and  outputs ($\Sigma_o$), (v)
  $E\subseteq L\times\L(X)\times\Sigma\times 2^X \times L$ is a finite
  set of edges, (vi)  $R:L\rightarrow\bbbn^X$ assigns a rate vector to
  each location, and (viii) $I:L\rightarrow\U(X)$ assigns an invariant
  to each location.
\end{definition}
\noindent
The semantics of NPTAs is a timed labelled transition system whose
states are pairs $(\ell,\nu)\in L\times\rplus^X$ with $\nu\models
I(\ell)$, and whose transitions are either delay
$(\ell,\nu)\arrow{d}(\ell,\nu')$ with $d\in\rplus$ and
$\nu'=\nu+R(\ell)\cdot d$, or discrete
$(\ell,\nu)\arrow{a}(\ell',\nu')$ if there is an edge
$(\ell,g,a,Y,\ell')$ such that $\nu\models g$ and $\nu'=\nu[Y]$. We
write $(\ell,\nu)\leadsto(\ell',\nu')$ if there is a finite
sequence of delay and discrete transitions from $(\ell,\nu)$
to $(\ell',\nu')$.

\paragraph{\bf{Networks of Priced Timed Automata}}
Following the compositional specification theory for timed systems in
\cite{hscc2010}, we shall assume that NPTAs are: (1)[{\sl
  Input-enabled}:] for all states $(\ell,\nu)$ and input actions
$\iota\in\Sigma_i$, $(\ell,\nu)\arrow{\iota}$, and (2) [{\sl
  Deterministic}:] for all states $(\ell,\nu)$ and actions
$a\in\Sigma$, whenever $(\ell,\nu)\arrow{a}(\ell',\nu')$ and
$(\ell,\nu)\arrow{a}(\ell'',\nu'')$ then $\ell'=\ell''$ and
$\nu'=\nu''$.

Whenever ${\cal A}^j=(L^j,X^j,\Sigma^j,E^j,R^j,I^j)$ ($j=1\ldots n$)
are NPTA, they are \emph{composable} into a \emph{closed network} iff
their clock sets are disjoint ($X^j\cap X^k=\emptyset$ when $j\neq
k$), they have the same action set ($\Sigma=\Sigma^j=\Sigma^k$ for all
$j,k$), and their output action-sets provide a partition of $\Sigma$
($\Sigma^j_o\cap\Sigma^k_o=\emptyset$ for $j\neq k$, and
$\Sigma=\cup_j \Sigma^j_o$).  For $a\in\Sigma$ we denote by $c(a)$ the
unique $j$ with $a\in\Sigma^j$.

\begin{definition}
  Let  ${\cal  A}^j=(L^j,X^j,\Sigma,E^j,R^j,I^j)$  ($j=1\ldots n$)  be
  composable    NPTAs.     Then    the   \emph{composition}    $({\cal
    A}_1\,|\ldots|\,{\cal   A}_n)$  is  the   NPTA  $\boldsymbol{{\cal
      A}}=(L,X,\Sigma,E,R,L)$   where    (i)   $L=\times_jL^j$,   (ii)
  $X=\cup_j X^j$, (iii) $R(\boldsymbol{\ell})(x)=R^j(\ell^j )(x)$ when
  $x\in  X^j$, (iv)  $I(\boldsymbol{\ell})=\cap_j I(\ell^j)$,  and (v)
  $(\boldsymbol{\ell},\cap_j  g_j,a,\cup_j  r_j,\boldsymbol{\ell'})\in
  E$ whenever $(\ell_j,g_j,a,r_j,\ell'_j)\in E^j$ for $j=1\ldots n$.
\end{definition}


\begin{figure}[htb]
  \centering
  \vspace*{-5mm}
  \begin{tabular}{*{5}{c}}
    \includegraphics[height=0.11\textheight]{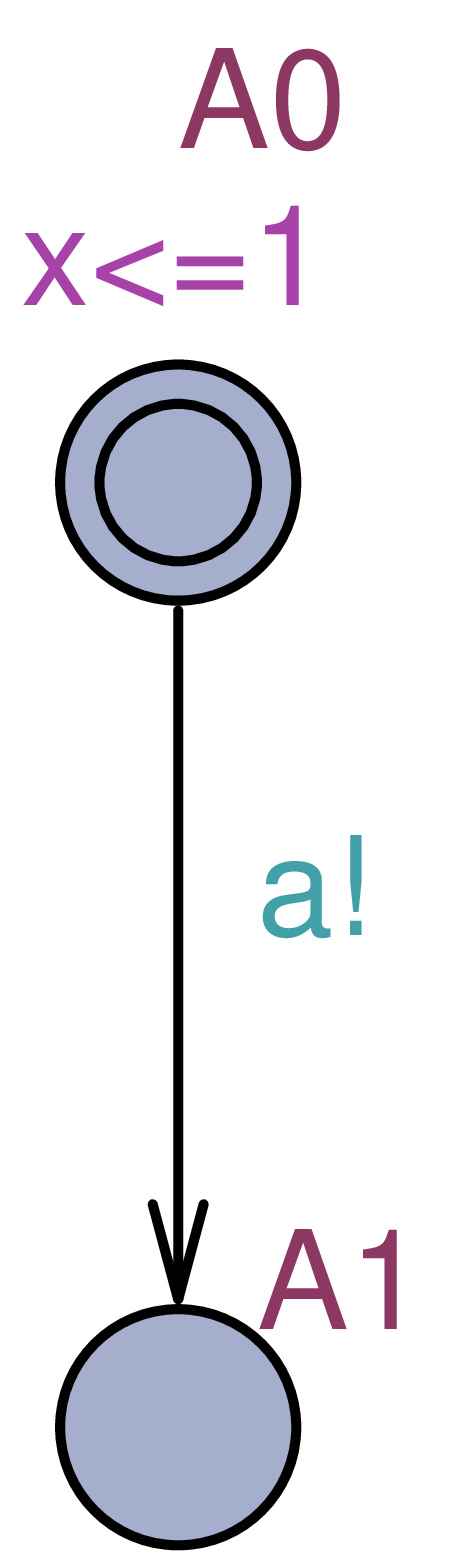} &
    \includegraphics[height=0.11\textheight]{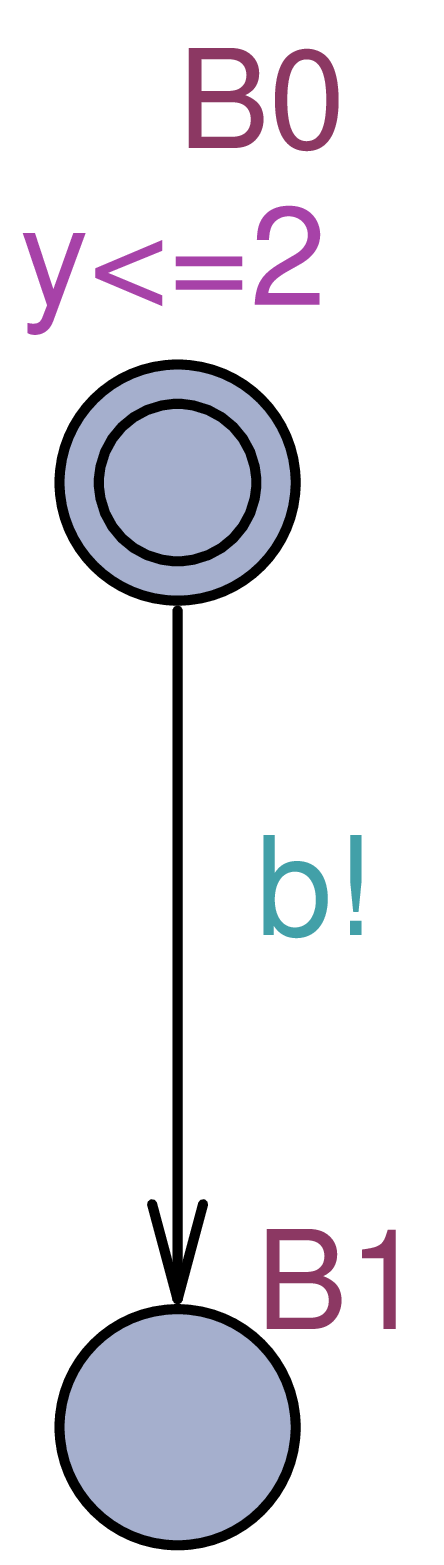} &
    \includegraphics[height=0.11\textheight]{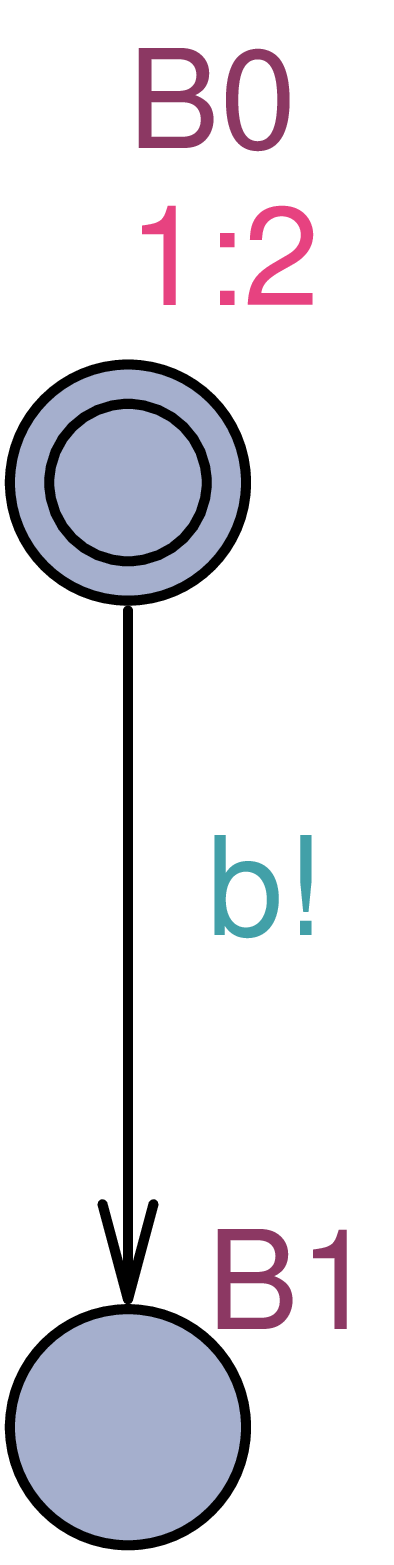} &
    \includegraphics[height=0.11\textheight]{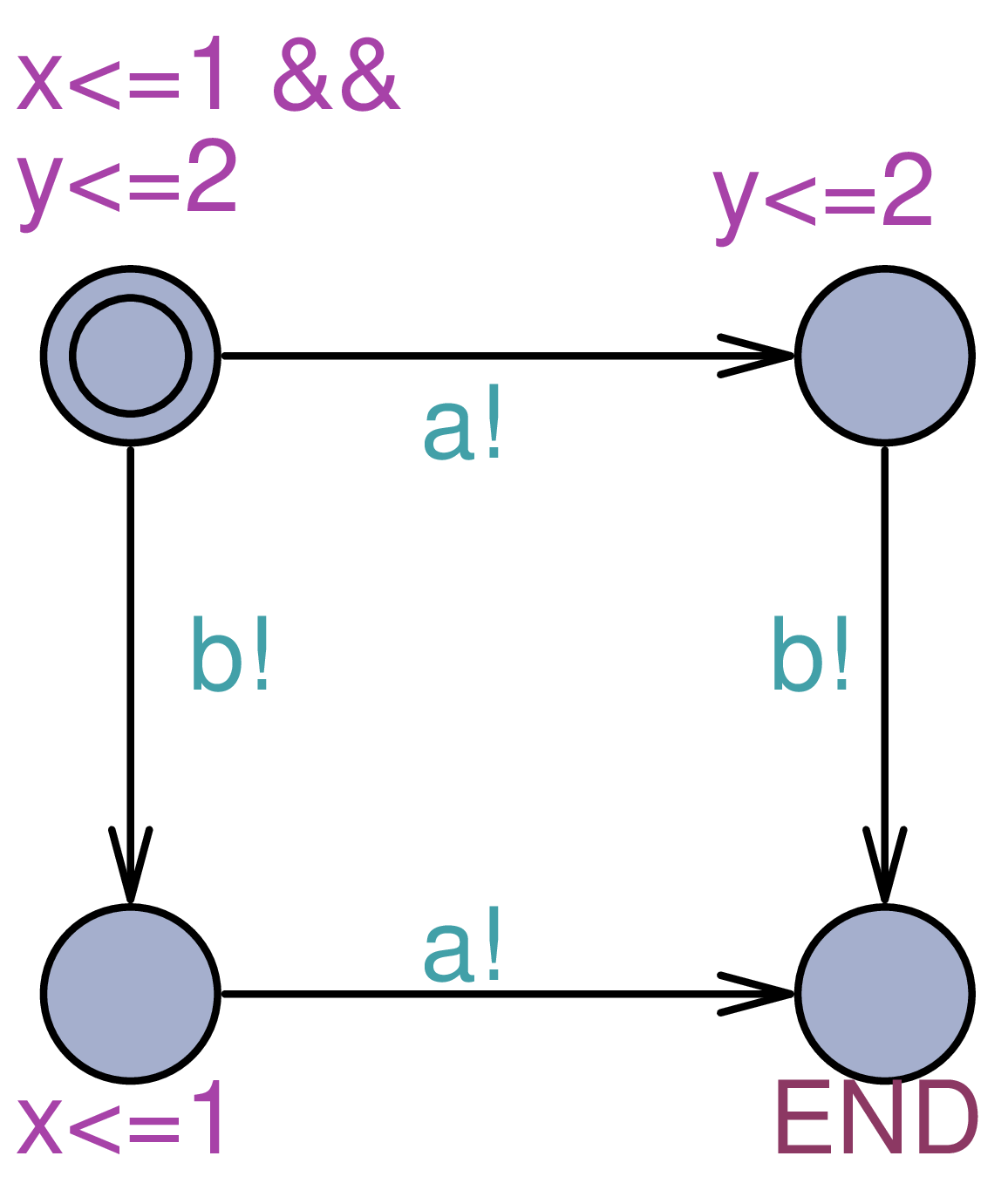} &
    \includegraphics[height=0.11\textheight]{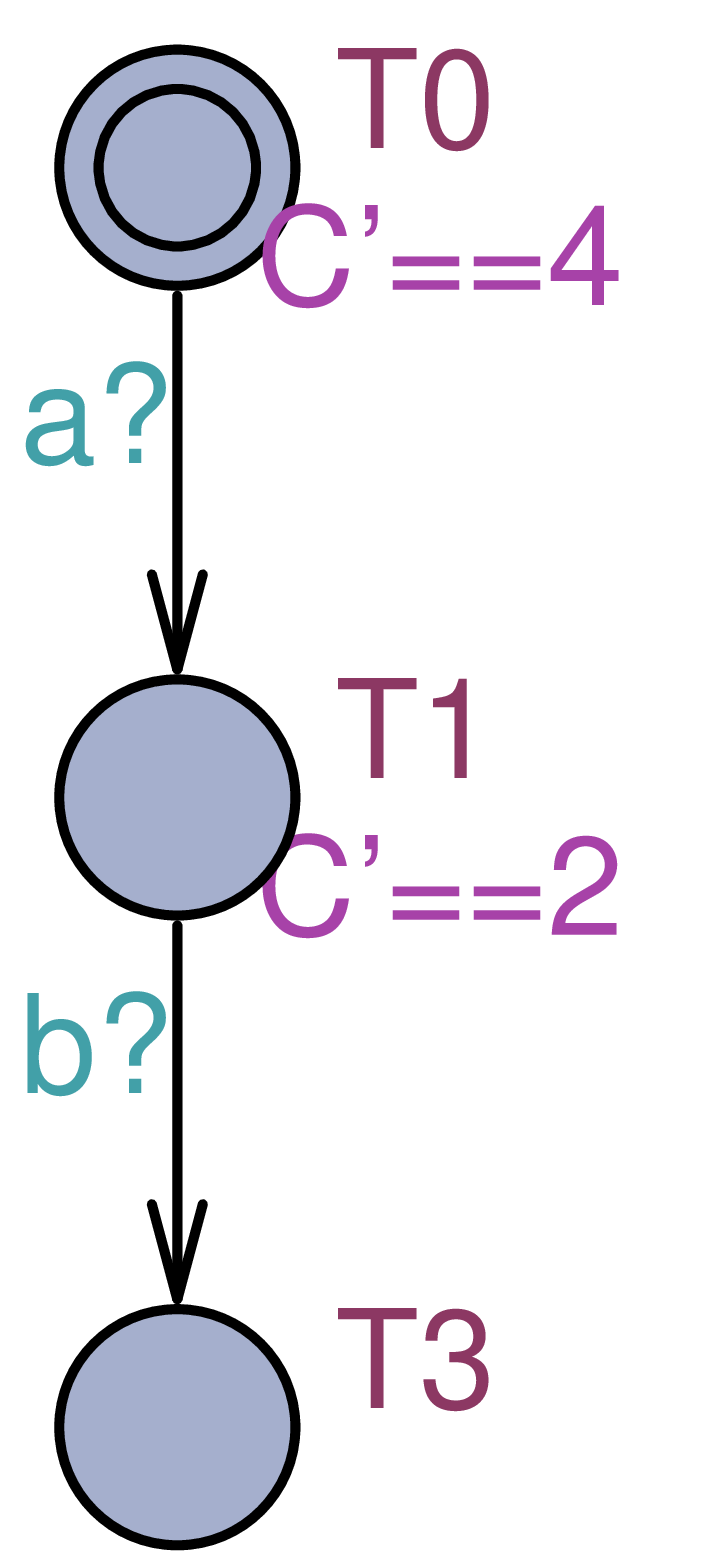} \\
    $A$ & $B$ & $B_r$ & $AB$ & $T$ \\
  \end{tabular}
  \vspace*{-2mm}
  \caption{Four composable NPTAs: $A, B$ and $T$; $A, B_r$ and $T$; and $AB$~and~$T$.}
  \label{fig:NPTA}
\end{figure}

\noindent
\textit{Example 1.}  Let $A$, $B$, $T$ and $AB$ be the priced timed
automata depicted in Fig.~\ref{fig:NPTA}\footnote{it is assumed that
  all components are completed with looping input transitions, where
  these are missing.}.  Then $A, B$ and $T$ are composable as well as
$AB$ and $T$. In fact the composite systems $(A|B|T)$ and $(AB|T)$ are
timed (and priced) bisimilar, both having the transition sequence:\smallskip \\
%
{\small
$
\big((A_0,B_o,T_0),[x=0,y=0,C=0]\big) \arrow{1}\arrow{a!}\\
\big((A_1,B_0,T_1),[x=1,y=1,C=4]\big) \arrow{1} \arrow{b!}\\
\big((A_1,B_1,T_2),[x=2,y=2,C=6]\big),$}\smallskip\\
demonstrating that the final location $T_3$ of $T$ is reachable with cost
$6$.


\section{Probabilistic Semantics of NPTA}
\label{sec:markov}

\renewcommand{\P}{\ensuremath{{\mathbb P}}\xspace}

Continuing Example 1
we may  realise that location $T_3$ of
the component $T$ is reachable within cost $0$ to $6$ and within total
time $0$ and $2$ in both $(A|B|T)$ and $(AB|T)$ depending on when (and
in  which order)  $A$ and  $B$ ($AB$)  chooses to  perform  the output
actions $a!$ and $b!$.  Assuming  that the choice of these time-delays
is  governed by probability  distributions, we  will in  this section
define a probability measure over sets of infinite runs of networks of
NPTAs. 

In contrast to the probabilistic semantics of timed automata in
\cite{BBBM08,BBBBG07} our semantics deals with networks and thus with
races between components.  Let ${\cal
  A}^j=(L^j,X^j,\Sigma,E^j,R^j,I^j)$ ($j=1\ldots n$) be a collection
of composable NPTAs.  Under the assumption of input-enabledness,
disjointness of clock sets and output actions, states of the the
composite NPTA $\boldsymbol{{\cal A}}=({\cal A}_1\,|\ldots|\,{\cal
  A}_n)$ may be seen as tuples ${\bf s}= (s_1,\ldots,s_n)$ where $s_j$
is a state of ${\cal A}^j$, i.e.  of the form $(\ell,\nu)$ where
$\ell\in L^j$ and $\nu\in\rplus^{X^j}$.  Our probabilistic semantics
is based on the principle of independency between components.
Repeatedly each component decides on its own -- based on a given delay
density function and output probability function -- how much to delay
before outputting and what output to broadcast at that moment.
Obviously, in such a race between components the outcome will be
determined by the component that has chosen to output after the
minimum delay: the output is broadcast and all other components may
consequently change state.

\paragraph{{\bf Probabilistic Semantics of NPTA Components}}

Let us first consider a  component ${\cal A}^j$ and let $\St^j$ denote
the  corresponding set of  states.  For  each state  $s=(\ell,\nu)$ of
${\cal  A}^j$  we shall  provide  probability  distributions for  both
delays and outputs.

The \emph{delay density function} $\mu_s$ over delays in $\rplus$ will
be either a uniform or an exponential distribution depending on the
invariant of $\ell$. Denote by $E_{\ell}$ the disjunction of guards
$g$ such that $(\ell,g,o,-,-)\in E^j$ for some output $o$. Denote by
$d(\ell,\nu)$ the infimum delay before enabling an output, i.e.
$d(\ell,\nu)=\inf\{d\in\rplus\,:\, \nu+R^j\cdot d \models E_{\ell}\}$,
and denote by $D(\ell,\nu)$ the supremum delay, i.e.
$D(\ell,\nu)=\sup\{d\in\rplus\,:\, \nu+R^j\cdot d \models
I^j(\ell)\}$.  If $D(\ell,\nu)<\infty$ then the delay density function
$\mu_s$ is a uniform distribution on $[d(\ell,\nu),D(\ell,\nu)]$.
Otherwise -- that is $I^j(\ell)$ does not put an upper bound on the
possible delays out of $s$ -- the delay density function $\mu_s$ is an
exponential distribution with a rate $P(\ell)$, where
$P:L^j\rightarrow\rplus$ is an \emph{additional} distribution rate
component added to the NPTA ${\cal A}^j$. For every state
$s=(\ell,\nu)$, the \emph{output probability function} $\gamma_s$ over
$\Sigma^j_o$ is the uniform distribution over the set $\{
o\,:\,(\ell,g,o,-,-)\in E^j\wedge \nu\models g\}$ whenever this set is
non-empty \footnote{otherwise a specific weight distribution can be 
specified and used instead.}.  We denote by $s^o$ the state after 
the output of $o$. Similarly, for every state $s$ and any input 
action $\iota$, we denote by $s^{\iota}$ the state after having 
received the input $\iota$.

\paragraph{{\bf Probabilistic Semantics of Networks of NPTA}}

We shall now see that while the stochastic semantics of each PTA is
rather simple (but quite realistic), arbitrarily complex stochastic
behavior can be obtained by their composition. 

Reconsider the closed network $\boldsymbol{{\cal A}}=({\cal
  A}_1\,|\ldots|\,{\cal A}_n)$ with a state space
$\St=\St_1\times\cdots\times\St_n$.  For ${\bf s}=(s_1,
\ldots,s_n)\in\St$ and $a_1a_2\ldots a_k\in\Sigma^{*}$ we denote by
$\pi({\bf s},a_1a_2\ldots a_k)$ the set of all maximal runs from ${\bf
  s}$ with a prefix $t_1a_1t_2a_2\ldots t_ka_k$ for some
$t_1,\ldots,t_n\in\rplus$, that is runs where the $i$'th action $a_i$
has been outputted by the component $A_{c(a_i)}$.  We now inductively
define the following measure for such sets of runs:

\begin{eqnarray*}
\lefteqn{\P_{\A}\big(\pi({\bf s},\,a_1a_2\ldots a_n)\big) = } \\
& \int_{t\geq 0}
    \mu_{s_c}(t) \cdot 
    \big(\prod_{j\not=c}\int_{\tau>t} \mu_{s_j}(\tau)\,d\tau \big)
    \cdot 
   \gamma_{{s_c}^t}(a_1) \cdot\\
&    \P_{\A}\big(\pi({\bf s}^t)^{a_1},\, a_2\ldots a_n)\big) \,\,dt
\end{eqnarray*}

\noindent
where   $c=c(a_1)$,    and   as    base   case   we    take   $P_{\A}(\pi({\bf s}),\varepsilon)=1$.

This definition requires a few words of explanation: at
the outermost level we integrate over all possible initial delays $t$.
For a given delay $t$, the outputting component $c=c(a_1)$ will choose
to   make    the   broadcast   at    time   $t$   with    the   stated
density. Independently,  the other components  will choose to  a delay
amount, which -- in  order for $c$ to be the winner  -- must be larger
than $t$; hence  the product of the probabilities  that they each make
such a  choice. Having decided for  making the broadcast  at time $t$,
the probability of actually outputting $a_1$ is included.  Finally, in
the  global state  resulting from  all components  having  delayed $t$
time-units and changed state according to the broadcasted action $a_1$
the probability of runs  according to the remaining actions $a_2\ldots
a_n$ is taken into account.

\paragraph{{\bf Logical Properties}} Following \cite{panangaden}, the
measure $\P_{\A}$ may be extended in a standard and unique way to the
$\sigma$-algebra generated by the sets of runs (so-called cylinders)
$\pi({\bf s},a_1a_2\ldots a_n)$. As we shall see this will allow us to
give proper semantics to a range of probabilistic time- and
cost-constrained temporal properties.  Let ${\cal A}$ be a NPTA. Then
we consider the following non-nested PWCTL properties:
\[ \psi \,\,::=\,\, \P\big(\Diamond_{C\leq c}\phi\big) \sim p 
\,\,\,\,|\,\,\,\, \P\big(\Box_{C\leq c}\phi\big)\sim p \]
where  $C$   is  an   observer  clock  (of   ${\cal  A}$),   $\phi$  a
state-property (wrt.  ${\cal  A}$) , $\sim\in\{<,\leq,=,\geq,>\}$, and
$p\in [0,1]$. For the semantics let ${\cal A}^{*}$ be the modification
of ${\cal  A}$, where the  guard $C\leq c$  has been conjoined  to the
invariant      of       all      locations      and       an      edge
$(\ell,\phi,o_\phi,\emptyset,\ell)$  has  been   added  to  all  edges
$\ell$, where $o_\phi$ is a new output action. Then:
\[  {\cal A}\models \P\big(\Diamond_{C\leq c}\phi\big) \sim p
\,\,\,\,\mbox{iff}\,\, \,\,
\P_{\cal A^{*}}\Big(\bigcup_{\sigma\in\Sigma^{*}}\pi(s_0,\sigma
o_{\phi})\Big) \sim p
\]
which is  well-defined since  the $\sigma$-algebra on  which $\P_{\cal
  A^{*}}$  is defined  is  closed under  countable  unions and  finite
intersections.    To   complete    the   semantics,   we   note   that
$\P(\Box_{C\leq      c}\phi)\sim     p$      is      equivalent     to
$(1-p)\sim\P(\Diamond_{C\leq c}\neg\phi)$.

\begin{figure}[t]
  \centering

%
  \begin{tabular}{cc}
\hspace{-10pt}
    \includegraphics[width=0.49\linewidth]{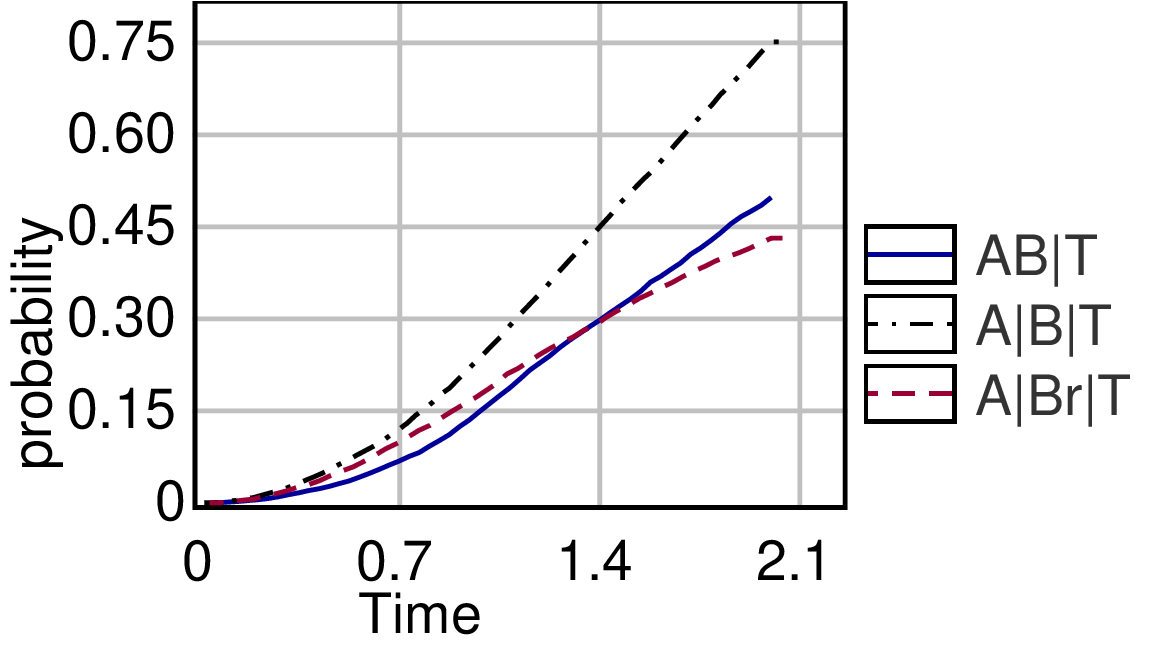} &
\hspace{-10pt}
    \includegraphics[width=0.49\linewidth]{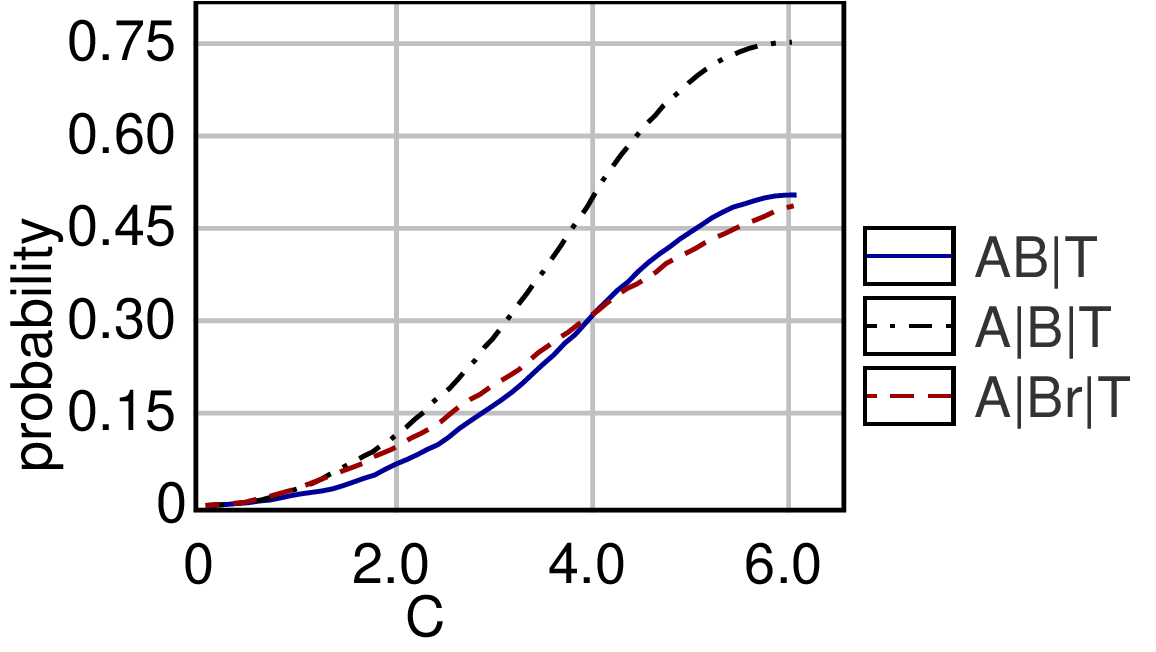} \\
    (a) & (b) 
  \end{tabular}
  \vspace*{-2mm}

%

  \caption{Cumulative probabilities for time and cost-bounded 
    reachability of $T_3$.}
  \label{CumTC:fig}
  \vspace*{-4mm}
\end{figure}

\begin{example}
  Reconsider  the Example  of Fig.~\ref{fig:NPTA}. Then  it  can be
  shown that  $(A|B|T)\models \P\big(\Diamond_ {t\leq 2}T_3\big)=0.75$
  and   $(A|B|T)\models   \P\big(\Diamond_  {C\leq   6}T_3\big)=0.75$,
  whereas $(AB|T)\models  \P\big(\Diamond_ {t\leq 2}T_3\big)=0.50$ and
  $(AB|T)\models     \P\big(\Diamond_     {C\leq     6}T_3\big)=0.50$.
  Fig.~\ref{CumTC:fig} gives  a time- and  cost-bounded reachability
  probabilities  for $(A|B|T)$  and $(AB|T)$  for a  range  of bounds.
  Thus, though  the two NPTAs  satisfy the same WCTL  properties, they
  are obviously quite different with  respect to PWCTL. The NPTA $B_r$
  of Fig.~\ref{fig:NPTA} is a variant of $B$, with the uniform delay 
  distribution enforced by
  the   invariant  $y\leq   2$  being   replaced  by   an  exponential
  distribution  with rate  $\frac{1}{2}$.  Here $(A|B_r|T)$  satisfies
  $\P\big(\Diamond_     {t\leq     2}T_3\big)\approx     0.41$     and
  $\P\big(\Diamond_ {C\leq 6}T_3\big)\approx 0.49$.
\end{example}

\section{Statistical Model Checking for NPTA}
\label{sec:smc1}
\def\Sys{\mathcal{S}}

As we pointed out, most of model checking problems for NPTAs and PWCTL
(including reachability)  are undecidable.  Our  solution is to  use a
technique that  approximates the answer.  We rely  on {\em Statistical
  Model  Checking}  (SMC)\cite{You05a,SVA04},  that  is  a  series  of
simulation-based techniques that generate runs of the systems, monitor
them, and  then use algorithms from  statistics to get  an estimate of
the entire  system.  At  the heart  of any SMC  approach, there  is an
algorithm used to  generate runs of the system  following a stochastic
semantics. We propose such an algorithm for NPTAs corresponding to the
stochastic semantics proposed  in Section~\ref{sec:markov}.   Then, we
recap existing  statistic algorithms, providing the basis  for a first
SMC algorithm for NPTAs.

\paragraph{\bf Generating Runs of NPTA}
SMC is used for properties that can be monitored on finite runs. Here,
we propose an  algorithm that given an NPTA generates  a random run up
to  a cost bound  $c$ (with  time bounds  being a  simple case)  of an
observer clock $C$.  A run of  a NPTA is a sequence of alternations of
states
$\boldsymbol{s_0}\xrightarrow{d_0}\boldsymbol{s_0'}\xrightarrow{o_0}\boldsymbol{s_1}\xrightarrow{d_1}\dots\boldsymbol{s}_n$  obtained by  performing delays  $d_i$  and emitting
outputs $o_i$.  Here we consider  a network of NPTAs with states being
of  the  form  $(\boldsymbol{\ell},\nu)$.   We construct  random  runs
according to Algorithm~\ref{alg:runs}. We  start from an initial state
$(\boldsymbol{\ell_0},\nu_0)$   and   repeatedly  concatenate   random
successor states until  we reach the bound $c$  for the given observer
clock  $C$.   Recall  that $\nu(C)$  is  the  value  of $C$  in  state
$(\boldsymbol{\ell},\nu)$,   and   the  rate   of   $C$  in   location
$\boldsymbol{\ell}$ is $R(C)(\boldsymbol{\ell})$.  We use the notation
$\oplus$ to concatenate runs and  $tail(run)$ to access the last state
of a  run and $delay(\mu_s)$ returns  a random delay  according to the
delay     density     function     $\mu_s$     as     described     in
Section~\ref{sec:markov}.    The  statement   ``pick''   means  choose
uniformly among  the possible  choices.  The correctness  of Algorithm
\ref{alg:runs} with respect to the stochastic semantics of NPTAs given
in Section~\ref{sec:markov} follows from the Theorem below:

\begin{theorem}
  Let $\boldsymbol{\cal A}$ be a network of NPTAs. Then:
\[ \P\Big(
  RR _{\boldsymbol{\cal A}}\big((\boldsymbol{\ell_0},\nu_0),C,c\big)\models \Diamond_{C\leq c}\phi\Big) \,\, = \,\, \P_{\boldsymbol{\cal A}}\Big(\Diamond_{C\leq c}\,\phi\Big)
\]

\end{theorem}

\begin{algorithm}[t]
\caption{Random run for a NPTA-network $\boldsymbol{\cal A}$}
\label{alg:runs}
\begin{small}
\textbf{function} $RR _{\boldsymbol{\cal A}}((\boldsymbol{\ell_0},\nu_0),C,c)$\\
    \nl
    $run:=(\boldsymbol{\ell},\nu):=tail(run):=(\boldsymbol{\ell_0},\nu_0)$
    \\   
    \nl \While{$\nu(C)<c$}
    {
      \nl \textbf{for} {$i = 1$ to $|\boldsymbol{\ell}|$} \textbf{do}
           $d_i:=delay(\mu_{(\ell_i,\nu_i)})$\\
      \nl $d:=min_{1\le i\le |\boldsymbol{\ell}|}(d_i)$\\
      \nl \If {$d=+\infty\ \lor\ \nu(C)+d*R(\boldsymbol{\ell})(C)\ge c$}
      {
        \nl $d:=(c-\nu(C))/R(\boldsymbol{\ell})(C)$\\
        \nl \Return $run\oplus\xrightarrow{d}(\boldsymbol{\ell},\nu+d*R(\boldsymbol{\ell}))$\\
      }
      \nl \Else
      {
        \nl pick $k$ such that $d_k=d$; $\nu_d:=\nu+d*R(\boldsymbol{\ell})$\\
        \nl pick $\ell_k\xrightarrow{g,o,r}\ell_k'$ with $g(\nu_d)$ \\
        \nl
        $run:=run\oplus
        \xrightarrow{d}    (\boldsymbol{\ell},\nu_d)
        \xrightarrow{g,o,r}(\boldsymbol{\ell}[l_k'/l_k],[r\mapsto
        0](\nu_d))$\\

      }
    \nl $(\boldsymbol{\ell},\nu):=tail(run)$\\
    }
    \Return $run$
\end{small}
\end{algorithm}

\paragraph{\bf Statistical Model Checking Algorithms}
We  briefly  recap statistical  algorithms  permitting  to answer  the
following  two types of  questions :  (1) {\sl  Qualitative :}  Is the
probability  for a  given  NPTA $\boldsymbol{{\cal  A}}$  to satisfy  a
property  $\Diamond_{C\le  c}\phi$  greater  or  equal  to  a  certain
threshold  $\theta$  ?  and  (2)  {\sl  Quantitative  :} What  is  the
probability  for  $\boldsymbol{{\cal  A}}$ to  satisfy  $\Diamond_{C\le
c}\phi$.   Each run of  the system  is encoded  as a  Bernoulli random
variable  that is true  if the  run satisfies  the property  and false
otherwise.

\paragraph{\sl Qualitative Question.} This problem reduces to test the
hypothesis $H: p=\P_{\boldsymbol{{\cal A}}}(\Diamond_{C\le c}\phi) \ge \theta$ against
 $K: p < \theta$. To bound the
probability of making errors, we use strength parameters $\alpha$ and
$\beta$ and we test the hypothesis $H_0: p \ge p_0$ and $H_1: p \le
p_1$ with $p_0=\theta+\delta_0$ and $p_1=\theta-\delta_1$. The
interval $p_0-p_1$ defines an indifference region, and $p_0$ and $p_1$
are used as thresholds in the algorithm. The parameter $\alpha$ is the
probability of accepting $H_0$ when $H_1$ holds (false positives) and
the parameter $\beta$ is the probability of accepting $H_1$ when $H_0$
holds (false negatives). The above test can be solved by using Wald's
sequential hypothesis testing\,\cite{Wal04}. This test, which is
presented in Algorithm~\ref{alg:hypoTest1}, computes a proportion $r$
among those runs that satisfy the property. With probability 1, the
value of the proportion will eventually cross $\log(\beta/(1-\alpha)$
or $\log((1-\beta)/\alpha)$ and one of the two hypothesis will be
selected.

\begin{algorithm}[h]
\caption{Hypothesis testing\label{alg:hypoTest1}}
\begin{small}
\textbf{function} hypothesis($S$:model , $\psi$: property)\\
    \nl r:=0\\
    \nl \While{true}
    {
        \nl Observe the random variable $x$ corresponding to
        $\Diamond_{C\le c}\phi$ for a run.\\
        \nl $r := r + x*\log(p_1/p_0)+(1-x)*\log((1-p_1)/(1-p_0))$ \\
        \nl \textbf{if} $r \le \log(\beta/(1-\alpha))$ \textbf{then} accept $H_0$ \\
        \nl \textbf{if} $r \ge \log((1-\beta)/\alpha)$ \textbf{then} accept $H_1$
    }
\end{small}
\end{algorithm}


\paragraph{\bf Quantitative question}
This algorithm~\cite{vmcai10} computes the number $N$ of runs needed
in order to produce an approximation interval
${\lbrack}p-\epsilon,p+\epsilon{\rbrack}$ for $p=Pr(\psi)$ with a
confidence $1-\alpha$. The values of $\epsilon$ and $\alpha$ are
chosen by the user and $N$ relies on the Chernoff-Hoeffding bound as
shown in algorithm~\ref{alg:estimate}.

\begin{algorithm}[!htb]
\caption{Probability estimation\label{alg:estimate}}
\begin{small}
\textbf{function} estimate($S$:model , $\psi$: property, $\delta$: confidence, $\epsilon$: approximation)\\
    \nl $N := \ln(2/\alpha)/(2\epsilon^2)$, $a := 0$\\
    \nl \For{$i:=1$ to $N$}
    {
        \nl Observe the random variable $x$ corresponding to $\psi$ for a run.\\
        \nl $a := a+x$
    }
    \nl \Return $a/N$
\end{small}
\end{algorithm}

\section{Beyond ``Classical'' Statistical Model-Checking}
\label{sec:smc2}
\fxnote{Alex,Wang,Axel}


Here,    we    want     to    compare    $p_1=    \P_{\boldsymbol{\cal
    A}}(\Diamond_{C_1\le  c_1}\phi_1)$  and  $p_2=\P_{\boldsymbol{\cal
    A}}(\Diamond_{C_2\le  c_2}\phi_2)$  without  computing them,  with
clear applications  e.g.  in  determining the possible  improvement in
performance of a new control  program. In \cite{Wal04}, Wald has shown
that this  problem can be  reduced to a sequential  hypothesis testing
one. Our  contributions here  are (1) to  apply this algorithm  in the
formal  verification area,  (2) to  extend the  original  algorithm of
\cite{Wal04} to  handle cases where  we observe the same  outcomes for
both experiments, and  (3) to implement a parametric  extension of the
algorithm that allows to reuse  results on several timed bounds.  More
precisely, instead of comparing two probabilities with one common cost
bound  $C\le c$,  the new  extension does  it for  all the  $N$ bounds
$i*c/N$ with $i=1\dots N$ by reusing existing runs.

\paragraph{\bf Comparison Algorithm.} Let the efficiency of satisfying
$\Diamond_{C_1\le c_1}\phi_1$ over  runs be given by $k_1=p_1/(1-p_1)$
and  similarly   for  $\Diamond_{C_2\le  c_2}\phi_2$.    The  relative
superiority  of ``$\phi_2$ over  $\phi_1$'' is  measured by  the ratio
$u=\frac{k_2}{k_1}=\frac{p_2(1-p_1)}{p_1(1-p_2)}$.    If   $u=1$  both
properties are  equally good, if $u>1$, $\phi_2$  is better, otherwise
$\phi_1$ is better. Due to indifference region, we have two parameters
$u_0$ and  $u_1$ such  that $u_0<u_1$ to  make the decision.  If $u\le
u_0$  we favor  $\phi_1$  and if  $u\ge  u_1$ we  favor $\phi_2$.  The
parameter $\alpha$ is the probability of rejecting $\phi_1$ when $u\le
u_0$  and  the  parameter  $\beta$  is the  probability  of  rejecting
$\phi_2$ when $u\ge u_1$. An outcome for the comparison algorithm is a
pair         $(x_1,x_2)=(r_1\models\Diamond_{C_1\le        c_1}\phi_1,
r_2\models\Diamond_{C_2\le c_2}\phi_2)$ for two independent runs $r_1$
and   $r_2$.   In   Wald's   version  (lines   10--14   of   Algorithm
\ref{alg:com1}),  the outcomes  $(0,0)$ and  $(1,1)$ are  ignored. The
algorithm works  if it is guaranteed to  eventually generate different
outcomes. We extend the algorithm  with a qualitative test (lines 5--9
of Algorithm \ref{alg:com1}) to handle  the case when the outcomes are
always the same.
The      hypothesis     we      test      is     $\P_{\boldsymbol{\cal
    A}}(r_1\models\Diamond_{C_1\le             c_1}\phi_1            =
r_2\models\Diamond_{C_2\le c_2}\phi_2) \ge \theta$ for two independent
runs   $r_1$   and   $r_2$.   Typically   we   want   the   parameters
$p_0'=\theta+\delta_0$  (for the  corresponding hypothesis  $H_0$) and
$p_1'=\theta-\delta_1$ (for $H_1$) to be close to $1$.
Our version of the comparison algorithm is shown in
algorithm~\ref{alg:com1} with the following initializations:
$$\begin{array}{l}
a=\frac{\log(\frac{\beta}{1-\alpha})}{\log(u_1)-\log(u_0)},
 r=\frac{\log(\frac{1-\beta}{\alpha})}{\log(u_1)-\log(u_o)},
 c=\frac{\log(\frac{1+u_1}{1+u_0})}{\log(u_1)-\log(u_o)}
\end{array}$$


\begin{algorithm}[h]
\caption{Comparison of probabilities\label{alg:com1}}
\begin{small}
\textbf{function} comprise($S$:model , $\psi_1$, $\psi_2$: properties)\\
    \nl $check := 1$, $q := 0$, $t := 0$\\
    \nl \While{true}
    {
        \nl Observe the random variable $x_1$ corresponding to $\psi_1$ for a run. \\
        \nl Observe the random variable $x_2$ corresponding to $\psi_2$ for a run. \\
        \nl \If{check = 1}
        {
          \nl $x = (x_1 == x_2)$\\
          \nl $q = q + x*\log(p_1'/p_0')+(1-x)*\log((1-p_1')/(1-p_0'))$\\
          \nl \textbf{if} $q \le \log(\beta/(1-\alpha))$ \textbf{then return} \emph{indifferent}\\
          \nl \textbf{if} $r \ge \log((1-\beta)/\alpha)$ \text{then} $check=0$
        }
        \nl \If{$x_1 \neq x_2$}
        {
            \nl $a = a + c$,  $r = r + c$\\
            \nl \textbf{if} $x_1 = 0$ and $x_2 = 1$ \textbf{then} $t := t + 1$\\
            \nl \textbf{if} $t\le a$ \textbf{then} accept process $2$.\\
            \nl \textbf{if} $t\ge r$ \textbf{then} reject process $2$.
        }
    }
\end{small}
\end{algorithm}

\paragraph{\bf Parametrised Comparisons} We now generalise the comparison
algorithm to give answers not only for one cost bound $c$ but $N$ cost
bounds $i*c/N$ (with $i=1\dots N$). This algorithm is of particular
interest to generate distribution over timed bounds value of the
property. The idea is to reuse the runs of smaller bounds. When
$\Diamond_{C\le c}\phi_1$ or $\Diamond_{C\le c}\phi_2$ holds on some
run we keep track of the corresponding point in cost (otherwise the
cost value is irrelevant). Every pair or runs gives a pair of outcomes
$(x_1,x_2)$ at cost points $(c_1,c_2)$. For every $i=1\dots N$ we define the
new pair of outcomes $(y_{i_1},y_{i_2})=\big(x_1 \land (i\cdot c/N \ge
t_1\cdot rate_C),x_2 \land (i\cdot c/N \ge t_2\cdot rate_C)\big)$ for which we use our
comparison algorithm. We terminate the algorithm when a result for
every $i^{th}$ bound is known.

Let $a,r,c$ be the parameters of the previous comparison algorithm.
Let $a',r',c'$ be the parameters of the qualitative check of Section~\ref{sec:smc1}.
The procedure is shown in Algorithm~\ref{alg:com2}:

\begin{algorithm}[!htb]
\caption{The algorithm for parametrised probabilities comparison\label{alg:com2}}
\begin{small}
\textbf{function} comprise2($S$:model , $\phi_1$, $\phi_2$:
properties, C: clock, c: cost bound, N: \# of time intervals)\\
    \nl \For{$i:=1$ to $N$}
    {
        \nl $q_i := 0$, $a_i' := a'$, $r_i' := r'$, $t_i := 0$, $a_i := a$, $r_i := r$
    }
    \nl \Repeat{$stop = 1$}
    {
        \nl Observe $x_1$ corresponding to $\phi_1$ for a run at time $t_1$. \\
        \nl Observe $x_2$ corresponding to $\phi_2$ for a run at time $t_2$. \\
        \nl $stop := 1$ \\

        \nl \For{$i:=1$ to $N$}
        {
            \nl $y_1:=x_1 \land i*c/N \ge t_1*rate_C$ \\
            \nl $y_2:=x_2 \land i*c/N \ge t_2*rate_C$ \\
            
            \nl \If{$result_i = -2$}
            {
                \nl $a_i' := a_i' + c'$, $r_i' = r_i' + c'$ \\
                \nl \textbf{if} $y_1 = y_2$ \textbf{then} $q_i := q_i + 1$ \\
                \nl \textbf{if} $q_i \le a_i'$ \textbf{then} $result_i := 0.5$ \\
                \nl \textbf{if} $q_i \ge r_i'$ \textbf{then} $result_i := -1$
            }
            
            \nl \If{$result_i < 0$ and $y_1 \neq y_2$}
            {
                \nl  $a_i := a_i + c$, $r_i = r_i + c$\\
                \nl \If{$y_1 = 0$ and $y_2 = 1$}
                {
                    \nl $t_i := t_i + 1$
                }
                \nl \textbf{if} $t_i\le a_i$ \textbf{then} $result_i := 1$ \\
                \nl \textbf{if} $t_i\ge r_i$ \textbf{then} $result_i := 0$
            }

            \nl \textbf{if} $result_i < 0$ \textbf{then} $stop := 0$.
        }
    }
\end{small}
\end{algorithm}

The results for every $i^{th}$ bound are three-valued: $0$ means
$\phi_2$ is rejected, $1$ means $\phi_2$ is accepted, and $0.5$
means indifference.

\section{Case Studies}
\label{sec:casestudies}

We  have  extended  \uppaal  with  the algorithms  described  in  this
paper. The implementation provides access to all the powerful features
of the  tool, including user defined  functions and types,  and use of
expressions in guards, invariants, clock-rates as well as delay-rates.
Also  the  implementation   supports  branching  edges  with  discrete
probabilities  (using weights),  thus  supporting probabilistic  timed
automata (a feature for which  our stochastic semantics of NPTA may be
easily extended). Besides  these additional features, the case-studies
reported  below (as  well as  the plots  in the  previous part  of the
paper) illustrate the  nice features of the new plot composing GUI  
of the tool. 
For more    results   including models  of the case-studies see
\href{http://www.cs.aau.dk/~adavid/smc/}
{\it http://www.cs.aau.dk/\~{}adavid/smc/}.



\begin{figure}[htb]
\centering
\vspace*{-1mm}
a)\includegraphics[width=0.7\linewidth]{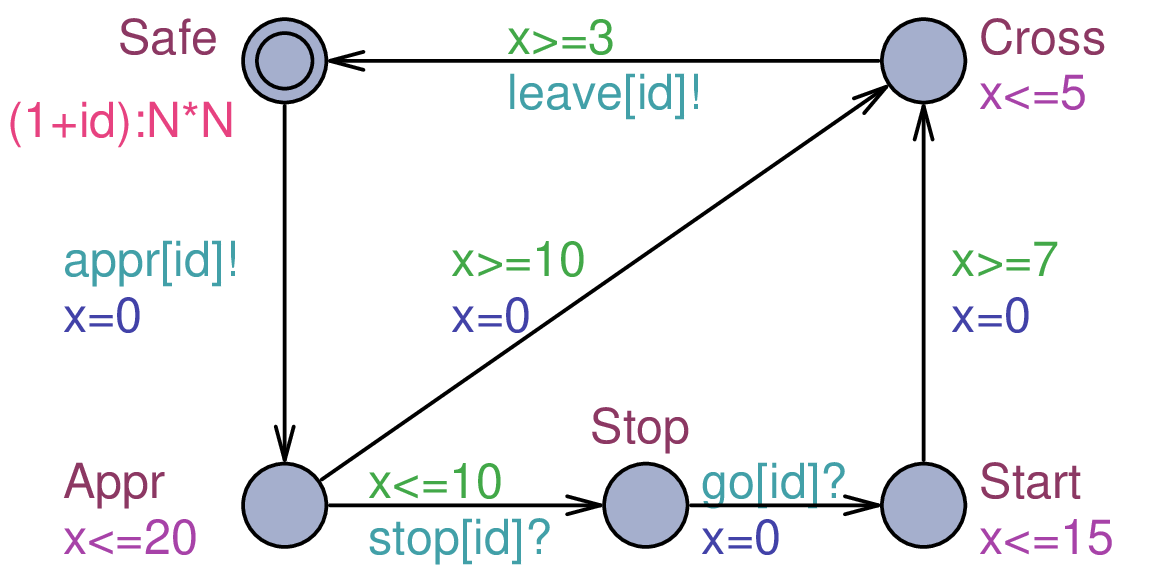} \medskip\\
b)\includegraphics[width=0.7\linewidth]{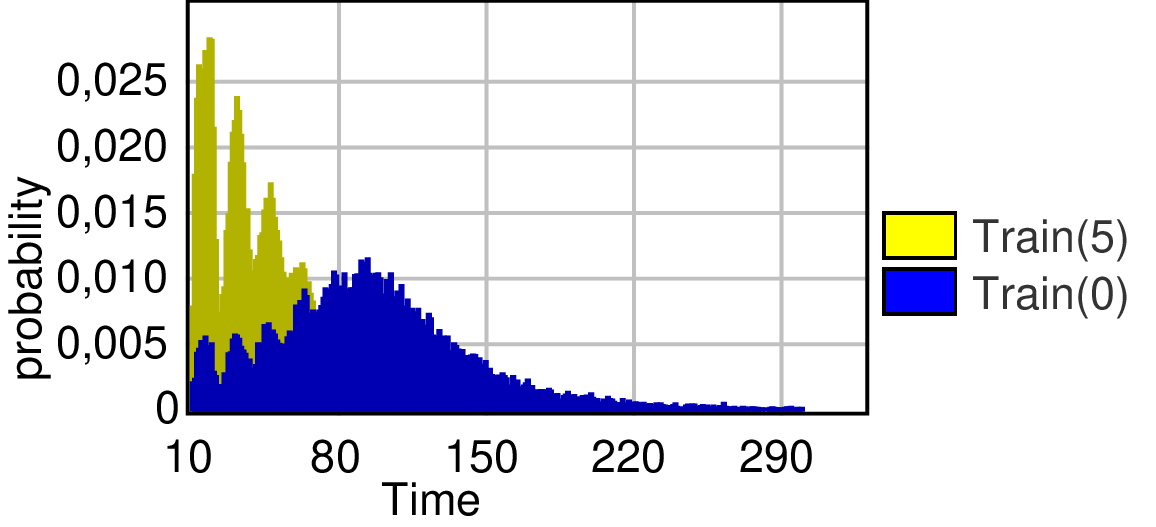} \\
\caption{Template of a train (a) and probability density distributions for
$\Diamond_{T\leq t} {\mathtt{Train(0).Cross}}$ and $\Diamond_{T\leq t} {\mathtt{Train(5).Cross}}$.
}
\label{fig:dist1}
\vspace*{-1mm}
\end{figure}

\paragraph{\bf Train-Gate Example}
We consider the train-gate example~\cite{DBLP:conf/sfm/BehrmannDL04},
where $N$ trains want to cross a one-track bridge. We extend the
original model by specifying an arrival rate for Train~$i$
($(i+1)/N$).
Trains are then approaching, but  they can be stopped before some time
threshold. When  a train  is stopped, it  can start  again. Eventually
trains cross the bridge and go  back to their safe state. The template
of  these  trains  is  given  in  Fig.~\ref{fig:dist1}(a).  Our  model
captures the  natural behavior of arrivals with  some exponential rate
and random delays chosen with uniform distributions in states labelled
with invariants.  The tool  is used to  estimate the  probability that
Train~$0$ and Train~$5$ will cross the bridge in less than $100$ units
of time.
Given  a  confidence  level  of  $0.05$  the  confidence  intervals
returned are  $[0.541,0.641]$ and  $[0.944,1]$. The tool  computes for
each time  bound $T$  the frequency  count of runs  of length  $T$ for
which   the   property   holds.  Figure~\ref{fig:dist1}(b)   shows   a
superposition of  both distributions  obtained directly with  our tool
that provides a plot composer for this purpose.

The distribution for Train~$5$ is the one with higher probability at the
beginning, which confirms that this train is indeed the faster one. An
interesting point is to note the valleys in the probability densities
that correspond to other trains conflicting for crossing the
bridge. They are particularly visible for Train~$0$. The number of
valleys corresponds to the number of trains. This is clearly not a
trivial distribution (not even uni-modal) that we could not have guessed manually even from
such a simple model. In addition, we use the qualitative check to
cheaply refine the result to $[0.541,0.59]$ and $[0.97,1]$.

%
\begin{figure}[htb]
\centering
\includegraphics[angle=-90,width=\linewidth]{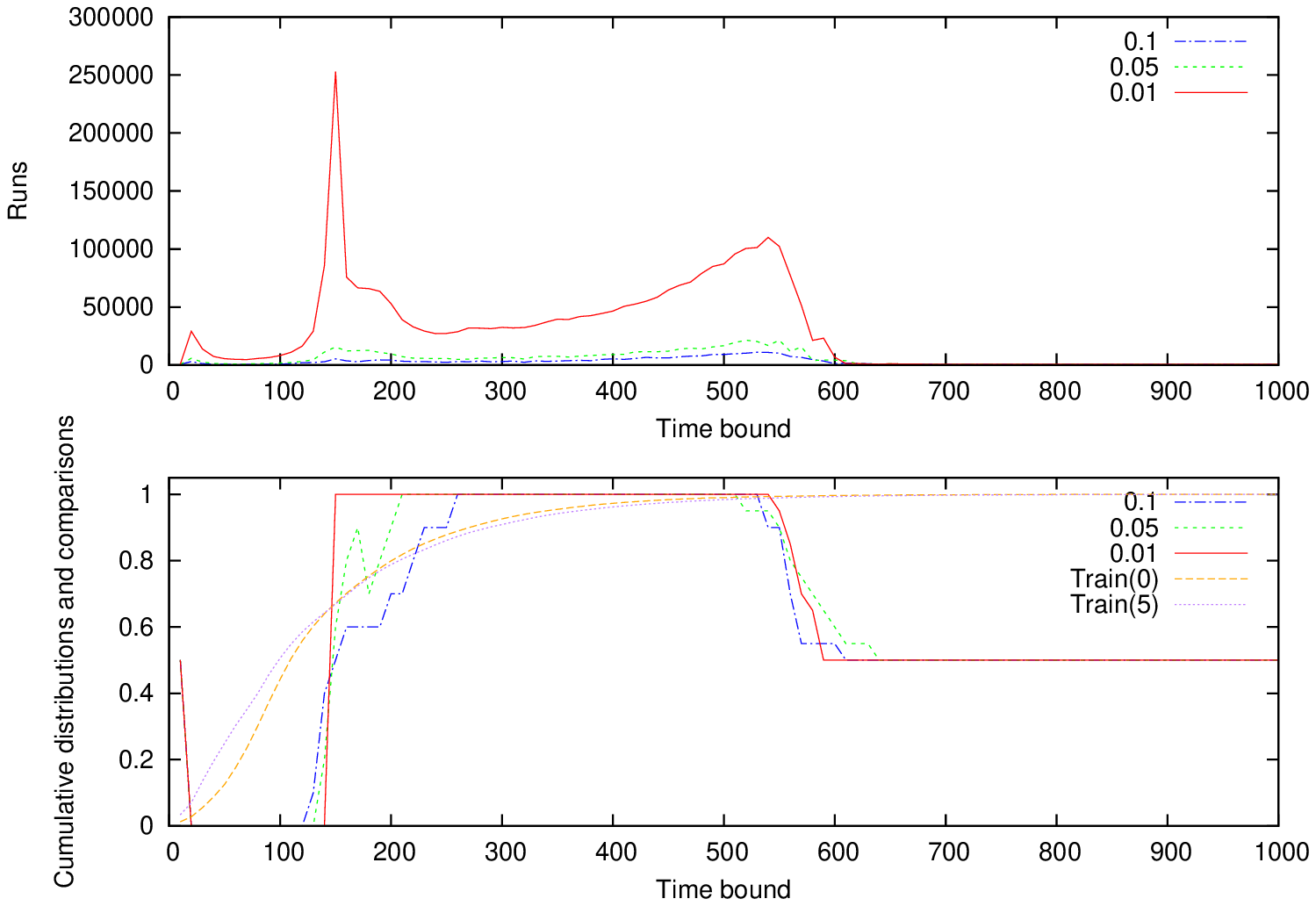}
\vspace*{-2mm}
\caption{Comparing trains $0$ and $5$.}
\label{fig:trains}
\vspace*{-3mm}
\end{figure}

We then compare the probability for Train~$0$ to cross when all other
trains are stopped with the same probability for Train~$5$.  In the
first plot (Fig.~\ref{fig:trains} top), we check the same property with $100$ different time
bounds from $10$ to $1000$ in steps of $10$ and we plot the number of
runs for each check.  These experiments only check for the specified
bound, they are not parametrised. In the second plot, we use the
parametric extension presented in Section~\ref{sec:smc2} with a
granularity of 10 time units. We configured the thresholds $u_0$ and
$u_1$ to differentiate the comparisons at $u_0=1-\epsilon$ and
$u_1=1+\epsilon$ with $\epsilon=0.1,0.05,0.01$ as shown on the
figure. In addition, we use a larger time bound to visualise the
behaviors after $600$ that are interesting for our checker. In the
first plot of Fig.~\ref{fig:trains}, we show for each time bound the
average of runs needed by the comparison algorithm repeated 30 times
for different values of $\epsilon$. In the bottom plot, we first
superpose the cumulative probability for both trains (curves Train~0
and Train~5) that we obtain by applying the quantitative algorithm of
Section~\ref{sec:smc1} for each time bound in the
sampling. Interestingly, before that point, train $5$ is better and
later train $0$ is better. Second, we compare these probabilities by
using the comparison algorithm (curves 0.1 0.05 0.01). This algorithm
can retrieve 3 values: 0 if Train~0 wins, 1 if Train~5 wins and 0.5
otherwise. We report for each time bound and each value of $\epsilon$ the
average of these values for 30 executions of the algorithm. 


\begin{table}[htb]
\vspace*{-1mm}
\centering
\caption{Sequential and parallel check comparison.}
\label{tab:comp}
\begin{tabular}{l@{\hspace{1em}}r@{\hspace{1em}}r@{\hspace{1em}}r}
\toprule
$\epsilon$ & 0.1 & 0.05 & 0.01 \\ \hline
sequential & 92s & 182s & 924s \\
parallel & 5s & 12s & 92s \\ 
\bottomrule
\end{tabular}
\end{table}
In addition, to evaluate the efficiency of computing all results at
once to obtain these curves, we measure the accumulated time to check
all the $100$ properties for the first plot (sequential check)%
, and the time to obtain all the results at once (parallel check)%
. The results are shown in Table~\ref{tab:comp}.
The experiments are done on a Pentium D at 2.4GHz and
consume very little memory. The parallel check is about 10 times
faster\footnote{The implementation checks simulations sequentially using 
a single thread.}. In fact it is limited by the highest number of runs 
required as shown by the second peak in Fig.~\ref{fig:trains}. 
The expensive part is to generate the runs so reusing them is important.


Note that at the beginning and at the end, our algorithm aborts the
comparison of the curves, which is visible as the number of runs
is sharply cut.

\paragraph{\bf Lightweight Media Access Control Protocol}

The Lightweight Media Access Control (LMAC) protocol is used in sensor
networks to  schedule communication  between nodes.  This  protocol is
targeted for  distributed self-configuration, collision  avoidance and
energy  efficiency.  In  this  study we  reproduce  the improved  
{\sc Uppaal} model  from \cite{FHM2007}   without    verification
optimisations,  parametrise with network  topology (ring  and chain),
add probabilistic weights (exponential and  uniform) over  discrete delay
decisions and  examine statistical properties which  were not possible
to check before.
Based on~\cite{Hoesel}, 
our node model consumes 21, 22, 2 and 1 power units per time unit when a node is sending, receiving, listening for messages or being idle respectively.


\begin{figure}[!htb]
  \vspace*{-2mm}
  \begin{center}
    \includegraphics[width=0.8\linewidth]{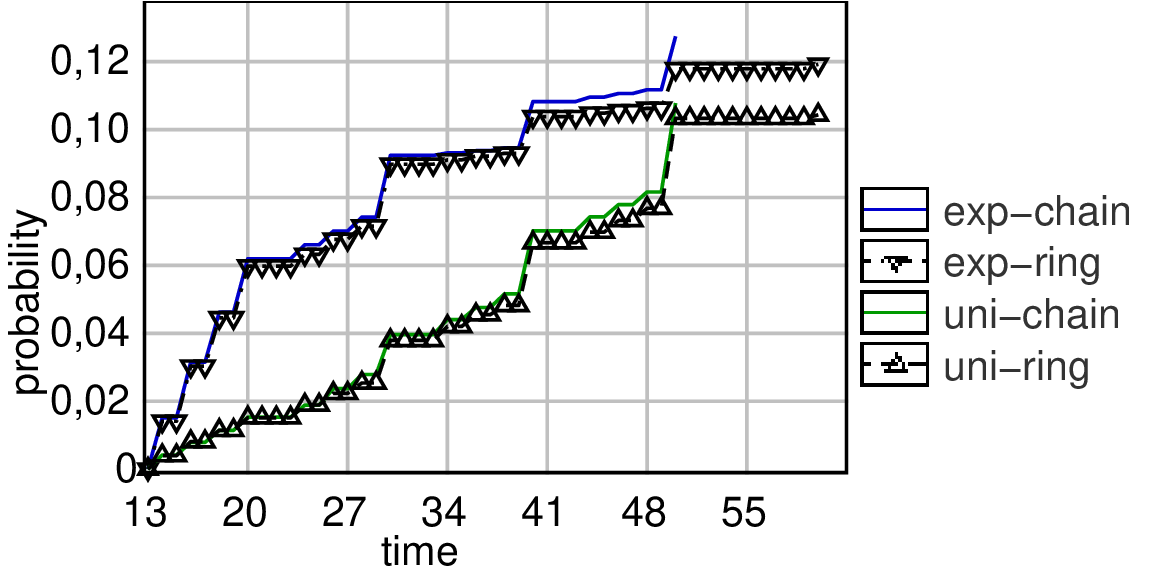}
  \end{center}
  \vspace*{-2mm}
  (a) Cumulative probability of collision over time.
  \begin{center}
    \includegraphics[width=0.8\linewidth]{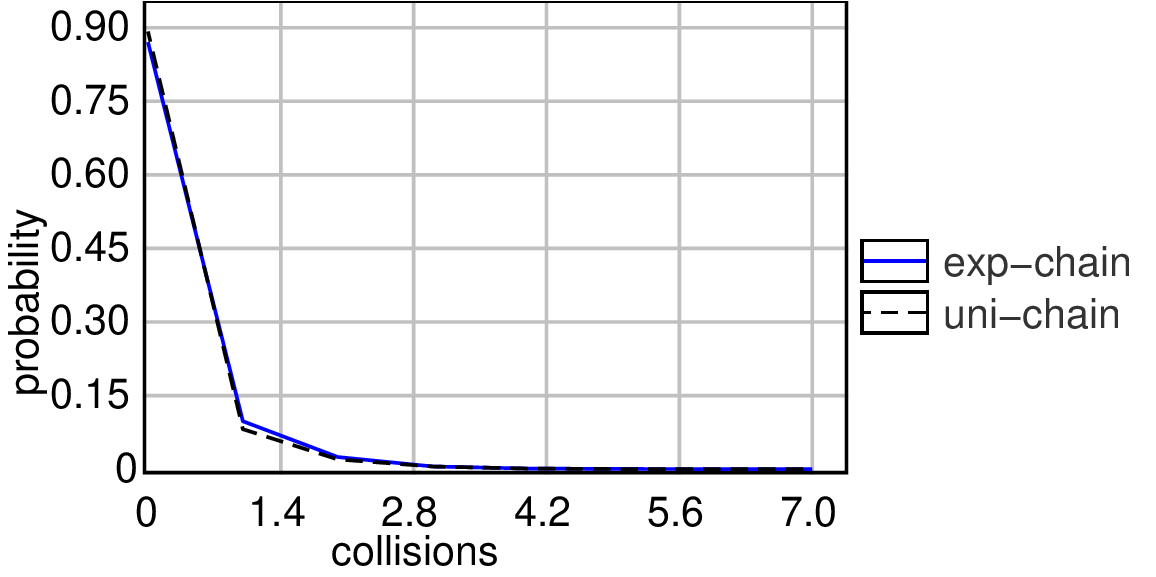}
  \end{center}
  \vspace*{-2mm}
  (b) Probability of having various numbers of collisions.
  \caption{Collision probabilities when using exponential and uniform weights in chain and ring topologies.}
  \label{fig:LMACc}
  \vspace*{-1mm}
\end{figure}
Fig.~\ref{fig:LMACc}a shows  that collisions  may happen in  all cases
and the  probability of collision is higher  with exponential decision
weights  than  uniform  decision  weights, but  seems  independent  of
topology (ring  or chain).  The probability of  collision stays stable
after 50 time units, despite  longer simulations, meaning that the
network may stay  collision free if the first  collisions are avoided.
We also  applied the method for  parametrised probability comparison
for the  collision probability.   The results are  that up to  14 time
units  the probabilities are  the same  and later  exponential weights
have higher  collision probability than uniform, but  the results were
inconclusive when comparing different topologies.

The  probable  collision counts  in  the  chain  topology are shown in
Fig.~\ref{fig:LMACc}b,  where  the  case   with  0  collisions  has  a
probability of 87.06\% and  89.21\% when using exponential and uniform
weights respectively.  The maximum number of probable collisions is 7
for both weight distributions despite very long  runs,  meaning
that the  network  eventually recovers  from collisions.

The probable  collision count in  the ring topology (not shown) yields
that there is no upper bound  of collision count as the collisions add
up indefinitely, but there is a fixed probability peak at 0 collisions
(87.06\% and 88.39\% using uniform and  exponential weights resp.) 
with a short tail up to 7 collisions (like in Fig.~\ref{fig:LMACc}b), 
long interval of 0 probability and then  small probability bump 
(0.35\% in total) at large  number  of  collisions.   Thus chances of
perpetual collisions are tiny.

\begin{wrapfigure}{r}{0.6\linewidth}
  \vspace*{-3.5mm}
  \includegraphics[width=\linewidth]{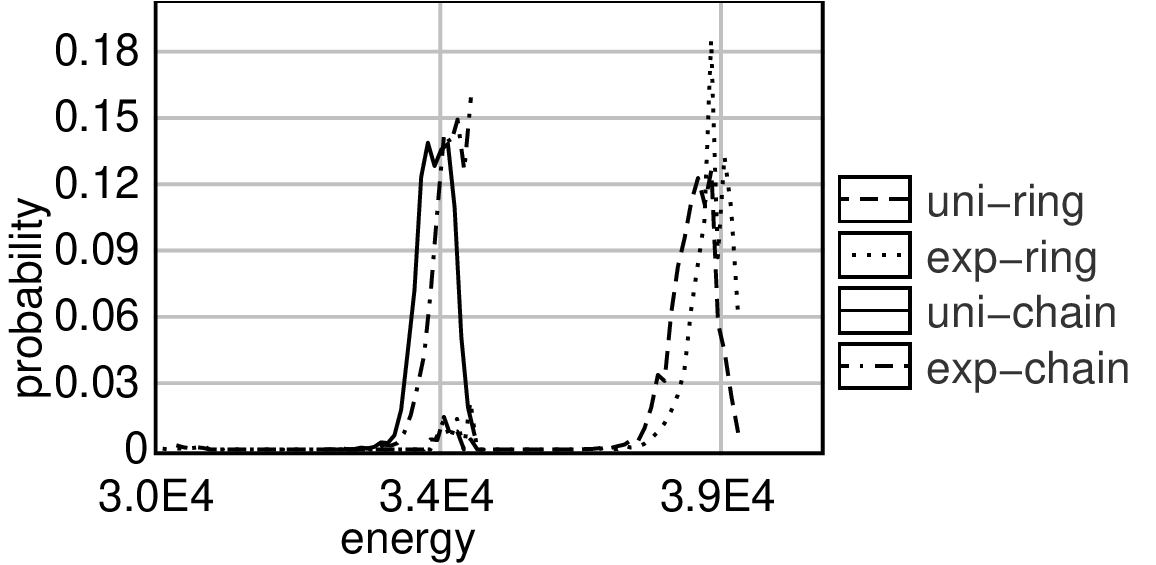}\\
  \vspace*{-6mm}
  \caption{Total energy consumption.}
  \label{fig:LMACe}
  \vspace*{-3mm}
\end{wrapfigure}
Fig.~\ref{fig:LMACe}  shows  energy  consumption probability  density:
using  uniform  and  exponential  weights   in  a  chain  and  a  ring
topologies.   Ring   topology  uses   more  power  (possibly   due  to
collisions),  and  uniform  weights   use  slightly  less  energy  than
exponential weights in these particular topologies.



\begin{figure}[htb]
\centering
\vspace*{-1mm}
\includegraphics[width=0.7\linewidth]{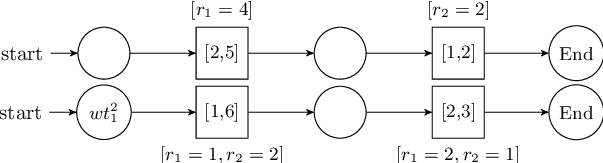}
\caption{Rectangles are busy states and circles are for waiting
when resources are not available. There are
$r_1=5$ and $r_2=3$ resources available.}\label{fig:DPAexample}
\end{figure}

\paragraph{\bf Duration Probabilistic Automata}
Duration  Probabilistic  Automata~\cite{MLK10} (DPA) are  used  for
modeling  job-shop problems.  A DPA  consists of  several  Simple DPAs
(SDPA). An SDPA is  a processing unit, a clock and a  list of tasks to
process sequentially.  Each task has an  associated duration interval,
from which its  duration is chosen (uniformly). Resources  are used to
model task races -- we allow  different resource  types  and different
quantities of each type. A fixed priority scheduler is used to resolve
conflicts. A DPA example is shown in Fig.~\ref{fig:DPAexample}.

\newcommand{\upp}{{\textsc Upp}}

DPA can be encoded in our tool (with a continuous or discrete time
semantics) or in \prism (discrete semantics), see the technical
report~\cite{PV11}.  In \prism, integer and boolean variables are used
to encode the current tasks and resources. \prism only supports the
discrete time model.  In \uppaal, a chain of waiting and task
locations is created for each SDPA. Guards and invariants
encode the duration of the task, and an array of integers contain the
available resources. The scheduler is encoded as a separate
template. We omit the resources and durations from the table for
simplicity, they are chosen arbitrarily for the experiment.
For \uppaal, both a discrete and continuous time versions
have been implemented. The performance of the translations is
measured on several case studies and shown in Tables~\ref{tab:durTestTab}
and~\ref{tab:nkTestTab}.  In the hypothesis testing column, \uppaal
({\upp}  in the table) uses the sequential hypothesis testing introduced
in Section~\ref{sec:smc1}, whereas \prism uses its own new implementation of
the hypothesis testing algorithm. In the estimation column, both
\uppaal and \prism use the quantitative check of
Section~\ref{sec:smc1}, but \uppaal is faster due to implementation
details.  For both tools, the error bounds used are
$\alpha=\beta=0.05$. In the hypothesis test, the indifference region
size is $0.01$, while we have $\epsilon=0.05$ for the quantitative
approach. The results show that \uppaal is faster than \prism even
with the discrete encoding, which currently is the only fair comparison.

\begin{table}[htbp]
  \centering
{\footnotesize
  \caption{Tool performance comparison.}
  \label{tab:durTestTab}
  \begin{tabular}{@{\extracolsep{-4pt}}*{3}{r}|*{3}{r}|*{3}{r}}
    \toprule
    \multicolumn{3}{c|}{\bf Parameters} &   \multicolumn{3}{c|}{\bf Estimation} & \multicolumn{3}{c}{\bf Hypothesis Testing} \\ 
    $n$ & $k$ & Duration & \prism &$\upp_d$ & $\upp_c$ & \prism &$\upp_d$ & $\upp_c$ \\ 
    \hline 
    10 & 10 & 4,8     & 42.2  & 8.7 & 6.9  & 64.1 & 1.0 & .3 \\ 
    10 & 10 & 8,16    & 60.3  & 11.3 & 7.2  & 49.4 & .7 & .3 \\ 
    10 & 10 & 16,32   & 91.8  & 13.4 & 7.0  & 77.1 & .9 & .4 \\ 
    10 & 10 & 32,64   & 126.0 & 14.8 & 7.0  & 65.8 & .9 & .3 \\ 
    10 & 10 & 64,128  & 176.8 & 16.3 & 7.0  & 83.4 & .9 & .3 \\ 
    \hline
    20 & 20 & 64,128 & - & 129.4 & 52.2 & - &5.2 & 1.6 \\ 
    20 & 20 & 128,256 & - & 146.4 & 52.1 & - & 8.1 & 1.8 \\ 
    20 & 20 & 256,512 & - & 173.8 & 52.3 & - & 11.6 & 1.8 \\ 
    \bottomrule
  \end{tabular}}
\end{table}

In the first test, we create a DPA with $n$ SDPAs, $k$ tasks per SDPA
and no resources. The duration interval of each task is changed and
the verification time is measured. In the second test, we choose
$n$, $k$ and let $m$ be the number of resource types. 
The resource usage and duration intervals are randomised.
The query  for the approximation  test is: ``What is the probability of
all SDPAs ending within $t$ time units?''. In the verification test, 
we ask  the  query:  ``Do  all SDPAs end within  $t$  time  units  with
probability greater than 40\%?''.  The value of  $t$ varies  for each
model as  it  was  computed  by  simulating the  system  369  times  and
represent the value for which at  least 60\% of the  runs reached the
final state.
\begin{table}[htbp]
  \centering
  {\footnotesize
  \caption{Comparison with various durations.}
  \label{tab:nkTestTab}
  \begin{tabular}{@{\extracolsep{-4.7pt}}*{3}{r}|*{3}{r}|*{3}{r}}
    \toprule
    \multicolumn{3}{c|}{\bf Parameters} &   \multicolumn{3}{c|}{\bf Estimation} & \multicolumn{3}{c}{\bf Hypothesis Testing} \\ 
    $n$ & $k$ & $m$ &\prism & $\upp_d$ & $\upp_c$ & \prism & $\upp_d$ & $\upp_c$ \\ 
    \hline
4 & 4 & 3 &  2.7 & 1.3 & 1.0 & 2.0 & .1 & .1 \\
6 & 6 & 3 &  7.7 & 3.4 & 2.6 & 3.9 & .2 & .3 \\
8 & 8 & 3 &  26.5 & 6.9 & 5.6 & 16.4 & .4 & .2 \\
    \hline
    20 & 40& 20 &  \multicolumn{3}{c|}{-} & $>$300 & 34.2 & 24.4 \\  
    30 & 40& 20 &   \multicolumn{3}{c|}{-}& $>$300 & 57.3 & 38.0 \\
    40 & 40& 20 &  \multicolumn{3}{c|}{-} & $>$300 & 67.4 & 70.0 \\ 
    40 & 20& 20 &  \multicolumn{3}{c|}{-} & $>$300 & 40.0 & 35.4 \\ 
    40 & 30 & 20&  \multicolumn{3}{c|}{-}  & $>$300 & 55.5 & 51.4\\ 
    \hline
    40 & 55 & 40& \multicolumn{5}{c|}{-}  & 219.5 \\ 
    50 & 55 & 40& \multicolumn{5}{c|}{-}  & 323.8 \\ 
    55 & 40 & 40& \multicolumn{5}{c|}{-}  & 307.0 \\ 
    55 & 50 & 40& \multicolumn{5}{c|}{-}  & 342.7 \\
    \bottomrule
  \end{tabular}}
\end{table}

\section{Conclusion and Future Work}
\label{sec:conclusion} 

This paper proposes a natural stochastic semantics for networks of
priced timed automata. The paper also explains how Statistical Model
Checking can be applied on the resulting model, handling case studies
that are beyond the scope of existing approaches.

The case studies show that models are more expressive, the tool is 
faster and capable of handling larger models than the scope of 
the state-of-the-art model-checker of stochastic systems.
The extended property language allows quantification of events with 
a limited impact in terms of probability and cost complementing 
critical property checks.
Hypothesis testing has an order of magnitude advantage in 
verification time over probability estimation, thus provides
an opportunity to gain leverage when more information is available.

There are many directions for future research. For example, 
the designer may have some prior knowledge about the probability 
of the property violation. 
This information could be used in a Bayesian fashion to
improve the efficiency of the test. If the system is assumed to be
``well-designed'', one can postulate that the property under
verification should rarely be falsified. In this case, the statistical
model checking algorithms will be efficient to compute the probability
of absence of errors.  Unfortunately, they will not be efficient to
compute the probability of making an error.  We propose to overcome
this problem by mixing existing SMC approaches with rare-event
techniques~\cite{Buc04}. Finally, it would also be of interest to
consider more elaborated properties~\cite{RP09,KZ09,YCZ10,BDDHP11} or
black-box systems~\cite{SVA04}.

\bibliographystyle{IEEEtran}
\bibliography{references,thesis,paper,s4-biblio}

\end{document}